\documentclass[twoside,11pt]{article}

\usepackage{latexsym}
\usepackage{amsmath}
\usepackage{amssymb}
\usepackage{amsthm}
\usepackage{epsfig}
\usepackage{graphics}
\usepackage{array}
\usepackage{rotating}
\usepackage{theapa}
\usepackage{jair}
\usepackage{algorithmic}
\usepackage{xspace}

\newlength{\figwidth}
\newlength{\maidwidth}
\newlength{\roadwidth}
\newlength{\smallfigwidth}
\newcommand{\reals}{{\rm I\!\hspace{-0.025em} R}}
\newcommand{\family}{\textit{Fam}}
\newcommand{\parents}{\textit{Pa}}
\newcommand{\adjoint}{\text{adj}}

\newcommand{\parass}{{\textbf{\itshape pa}}}

\newcommand{\clauses}{{\boldsymbol{C}}}
\newcommand{\bw}{{\boldsymbol{w}}}

\newcommand{\bu}{{\boldsymbol{u}}}
\newcommand{\bv}{{\boldsymbol{v}}}
\newcommand{\ba}{{\boldsymbol{a}}}

\newcommand{\bb}{{\boldsymbol{b}}}
\newcommand{\bt}{{\boldsymbol{t}}}
\newcommand{\bx}{{\boldsymbol{x}}}
\newcommand{\bX}{{\boldsymbol{X}}}
\newcommand{\bY}{{\boldsymbol{Y}}}
\newcommand{\bD}{{\boldsymbol{D}}}
\newcommand{\sF}{{\cal F}}
\newcommand{\sG}{{\cal G}}
\newcommand{\sD}{{\cal D}}
\newcommand{\sU}{{\cal U}}
\newcommand{\sX}{{\cal X}}
\newcommand{\sV}{{\cal V}}

\newcommand{\true}{\textit{true}\xspace}
\newcommand{\false}{\textit{false}\xspace}
\newcommand{\vars}{{\cal V}}

\newcommand{\C}{{\cal C}}
\newcommand{\bzero}{{\boldsymbol{0}}}
\newcommand{\commentout}[1]{}
\newcommand{\node}{y}

\newcommand{\bone}{{\boldsymbol{1}}}
\newcommand{\BN}{{\cal B}}

\renewcommand{\eqref}[1]{Equation~(\ref{#1})}
\newcommand{\figref}[1]{Figure~\ref{#1}}
\newcommand{\clmref}[1]{Claim~\ref{#1}}
\newcommand{\secref}[1]{Section~\ref{#1}}
\newcommand{\appref}[1]{Appendix~\ref{#1}}
\newcommand{\lemref}[1]{Lemma~\ref{#1}}
\newcommand{\thmref}[1]{Theorem~\ref{#1}}
\newcommand{\coderef}[1]{Step~\ref{#1}}
\newcommand{\exref}[1]{Example~\ref{#1}}
\newcommand{\subfigref}[2]{\figref{#1}(#2)}
\newcommand{\subfigrefn}[2]{\figref{#1}#2}
\newcommand{\subfigrefs}[2]{Figures~\ref{#1}(#2)}

\newcommand{\cont}{{\sf cont}\xspace}
\newcommand{\maidcont}{{\sf MAID cont}\xspace}
\newcommand{\efcont}{{\sf EF cont}\xspace}
\newcommand{\ipacont}{{\sf IPA+cont}\xspace}
\newcommand{\ipa}{{\sf IPA}\xspace}
\newcommand{\vk}{{\sf VK}\xspace}
\newcommand{\gambit}{{\sf Gambit}\xspace}
\newcommand{\GW}{GW\xspace}
\newcommand{\steps}{\textit{steps}}
\newcommand{\dom}{\textit{dom}}

\newtheorem{theorem}{Theorem}
\newtheorem{lemma}[theorem]{Lemma}
\newtheorem{example}[theorem]{Example}
\newtheorem{corollary}[theorem]{Corollary}
\newtheorem{claim}{Claim}[theorem]

\title{A Continuation Method for Nash Equilibria in Structured Games}

\author{Ben Blum \email{bblum@cs.berkeley.edu} \\
\addr  University of California, Berkeley \\
Department of Electrical Engineering and Computer Science \\
Berkeley, CA 94720
\AND Christian R. Shelton \email{cshelton@cs.ucr.edu} \\
\addr University of California, Riverside \\
Department of Computer Science and Engineering \\
Riverside, CA 92521
\AND Daphne Koller \email{koller@cs.stanford.edu} \\
\addr Stanford University \\
Department of Computer Science \\
Stanford, CA 94305
}

\begin{document}
\jairheading{25}{2006}{457-502}{11/05}{4/06}
\ShortHeadings{A Continuation Method for Nash Equilibria in Structured Games}{Blum, Shelton, \& Koller}
\firstpageno{457}

\maketitle
\begin{abstract}
Structured game representations have recently attracted interest as models for multi-agent artificial intelligence scenarios, with rational behavior most commonly 
characterized by Nash equilibria.  
This paper presents efficient, exact algorithms for computing Nash equilibria in 
structured game representations, including both graphical games and multi-agent influence
diagrams (MAIDs).  The algorithms are derived from a continuation
method for normal-form and extensive-form games due to Govindan and
Wilson; they follow a trajectory through a space of perturbed games and 
their equilibria, exploiting game structure
through fast computation of the Jacobian of the payoff function.
They are theoretically guaranteed to find at least one equilibrium of the
game, and may find more.  Our approach provides the first efficient algorithm
for computing exact equilibria in graphical games with arbitrary
topology, and the first algorithm to exploit fine-grained structural
properties of MAIDs. Experimental results are presented demonstrating the 
effectiveness of the algorithms and comparing them to predecessors.  The 
running time of the graphical game algorithm is similar to, and often
better than, the running time of previous approximate algorithms. The
algorithm for MAIDs can effectively solve games that are much larger than
those solvable by previous methods.
\end{abstract}

\section{Introduction}
\label{sec:intro}
In attempting to reason about interactions between multiple agents,
the artificial intelligence community has recently developed an
interest in game theory, a tool from economics.  Game theory is a very general
mathematical formalism for the representation of complex multi-agent
scenarios, called games, in which agents choose actions and then receive 
payoffs that depend on the outcome of the game.
A number of new game representations have been introduced in the past few years that
exploit structure to represent games more efficiently.
These representations are inspired by graphical models for probabilistic 
reasoning from the artificial intelligence literature, and include
\emph{graphical games} \cite{KeaLitSin01}, \emph{multi-agent influence
diagrams (MAIDs)} \cite{KolMil01}, \emph{G nets} \cite{LaM00}, and
\emph{action-graph games} \cite{BhaLey04}.

Our goal is to describe rational behavior in a game.  In game theory, a description of the behavior of all agents in the
game is referred to as a \emph{strategy profile}: a joint assignment
of strategies to each agent.  The most basic criterion to look for in a strategy profile is 
that it be optimal for each agent, taken individually: 
no agent should be able to improve
its utility by changing its strategy.  The fundamental
game theoretic notion of a \emph{Nash equilibrium}~\cite{Nas51}
satisfies this criterion precisely.  A Nash equilibrium is a strategy
profile in which no agent can improve its payoff by deviating
unilaterally --- changing its strategy while all other agents hold
theirs fixed.  There are other types of game theoretic solutions, but 
the Nash equilibrium is the most fundamental and is
often agreed to be
a minimum solution requirement.

Computing equilibria can be difficult for several reasons.
First, game representations themselves can grow quite large.
However,
many of the games that we would be interested in solving do not
require the full generality of description that leads to 
large representation size.  
The structured game representations 
introduced in AI
exploit structural properties
of games to represent them more compactly.  Typically, this structure
involves locality of interaction --- agents are only
concerned with the
behavior of a subset of other agents.

One would hope that more compact representations might lead to more
efficient computation of equilibria than would be possible with
standard game-theoretic solution algorithms (such as those described by 
\citeR{McKMcL96}).
Unfortunately, even with compact
representations, games are quite hard to solve; we present a result
showing that finding Nash equilibria beyond a single trivial one is
NP-hard in the types of structured games that we consider.

In this paper, we describe a set of algorithms for computing equilibria
in structured games that perform quite well,
empirically.  Our algorithms are in the family of \emph{continuation methods}.
They begin with a
solution of a trivial perturbed game, then track this solution as the
perturbation is incrementally undone, following a trajectory through
a space of equilibria of perturbed games until an equilibrium of the
original game is found.  Our algorithms are based on the recent work
of \citeauthor{GovWil02} \citeyear{GovWil02,GovWil03,GovWil04}
(\GW hereafter), which applies to standard game representations
(normal-form and extensive-form).  
The algorithms of \GW are of great
interest to the computational game theory community in their own right;
\shortciteA{NudWorShoLey04} have tested them 
against other leading algorithms and found them, in certain cases,
to be the most effective available.  However, as with all other algorithms
for unstructured games, they are infeasible for very large games.
We show how game structure can
be exploited to perform the key computational step of the algorithms
of \GW, and also give an alternative presentation of their work.

Our methods address both graphical games and
MAIDs.
Several recent papers have presented methods for finding equilibria in
graphical games.  Many of the proposed
algorithms~\cite{KeaLitSin01,LitKeaSin02,VicKol02,OrtKea03} have
focused on finding approximate equilibria, in which each agent may in
fact have a small incentive to deviate.  These sorts of algorithms
can be problematic: approximations must be crude for reasonable
running times, and there is no guarantee of an exact equilibrium in
the neighborhood of an approximate one.  Algorithms that find exact
equilibria have been restricted to a narrow class of
games~\cite{KeaLitSin01}.  We present the first efficient algorithm
for finding exact equilibria in graphical games of arbitrary
structure.  We present experimental results showing that the running
time of our algorithm is similar to, and often better than, the
running time of previous approximate algorithms.  Moreover, our
algorithm is capable of using approximate algorithms as starting
points for finding exact equilibria.

The literature for MAIDs is more limited.  The algorithm
of~\citeA{KolMil01} only takes advantage of certain coarse-grained
structure in MAIDs, and otherwise falls back on generating and solving
standard extensive-form games.  Methods for related types of
structured games~\cite{LaM00} are also limited to coarse-grained
structure, and are currently unimplemented.  Approximate approaches
for MAIDs~\cite{Vic02} come without implementation details or timing
results.  We provide the first exact algorithm that can take
advantage of the fine-grained structure of MAIDs.  We present
experimental results demonstrating that our algorithm can solve MAIDs
that are significantly outside the scope of previous methods.

\subsection{Outline and Guide to Background Material}
Our results require background in several distinct areas, including
game theory, continuation methods, representations of graphical games,
and representation and inference for Bayesian networks.  Clearly, it
is outside the scope of this paper to provide a detailed review of all
of these topics.  We have attempted to provide, for each of these
topics, sufficient background to allow our results to be understood.

We begin with an overview of
game theory in \secref{sec:gametheory}, describing strategy
representations and payoffs in both normal-form games (single-move games) 
and extensive-form games (games with multiple moves through time).  All concepts
utilized in this paper will be presented in this section, but a more thorough treatment
is available in the standard text by \citeA{FudTir91}.  
In \secref{sec:structrep} we introduce the two structured game
representations addressed in this paper: graphical games (derived from
normal-form games) and MAIDs (derived from extensive-form games).  
In \secref{sec:cplex} we give a result on the complexity of computing equilibria in both graphical games and MAIDs, with the proof deferred to \appref{nphardproof}.
We next outline continuation methods, the general scheme our algorithms use to
compute equilibria, in \secref{sec:contmeth}.  Continuation methods form a broad computational framework, and our presentation is therefore necessarily limited in scope; \citeA{Wat00} provides a more thorough grounding.  
In \secref{sec:contgnm}
we describe the particulars of applying continuation methods to
normal-form games and to extensive-form games.  The presentation is
new, but the methods are exactly those of \GW.  

In \secref{sec:exstruct}, we present our main contribution: exploiting
structure to perform the algorithms of \GW efficiently on both 
graphical games and MAIDs.  We show 
how Bayesian network inference in MAIDs can be
used to perform the key computational step of the \GW algorithm
efficiently, taking advantage of finer-grained structure than
previously possible.  
Our algorithm utilizes, as a subroutine, the clique tree inference
algorithm for Bayesian networks.  Although we do not present the
clique tree method in full, we describe the properties of the
method that allow it to be used within our algorithm; we also provide
enough detail to allow an implementation of our algorithm using
a standard clique tree package as a black box.  For a more
comprehensive introduction to inference in Bayesian networks,
we refer the reader to the reference by \citeA{Cowetal99}.  
In \secref{sec:res}, we present running-time
results for a variety of graphical games and MAIDs.  We conclude
in \secref{sec:conc}.

\section{Game Theory}
\label{sec:gametheory}
We begin by briefly reviewing concepts from game theory used in this
paper, referring to the text by
\citeA{FudTir91} for a good introduction.  We use the notation employed by~\GW. 
Those readers more familiar with game theory may wish to skip directly to the
table of notation in 
\appref{sec:tablenot}.

A game defines an interaction
between a set $N = \{n_1,n_2,\ldots,n_{|N|}\}$ of agents.  Each agent $n \in N$ has a set
$\Sigma_n$ of available \emph{strategies}, where a strategy determines
the agent's behavior in the game.  The precise definition of the set
$\Sigma_n$ depends on the game representation, as we discuss below.  A
\emph{strategy profile} $\sigma
=(\sigma_{n_1},\sigma_{n_2},\ldots,\sigma_{n_{|N|}}) \in \prod_{n \in
N} \Sigma_n$ defines a strategy $\sigma_n \in \Sigma_n$ for each agent
$n \in N$.  Given a strategy profile $\sigma$, the game defines an
expected payoff $G_n(\sigma)$ for each agent $n \in N$.  We use
$\Sigma_{-n}$ to refer to the set of all strategy profiles of agents
in $N\setminus\{n\}$ (agents other than $n$) and $\sigma_{-n}\in
\Sigma_{-n}$ to refer to one such profile; we generalize this
notation to $\Sigma_{-n,n'}$ for the set of strategy profiles of
all but two agents.  If $\sigma$ is a strategy profile, and
$\sigma_n' \in \Sigma_n$ is a strategy for agent $n$, then
$(\sigma_n',\sigma_{-n})$ is a new strategy profile in which $n$
deviates from $\sigma$ to play $\sigma_n'$, and all other agents act
according to $\sigma$.

A solution to a game is a prescription of a strategy profile for the
agents.  In this paper, we use Nash equilibria as our solution concept
--- strategy profiles in which no agent can profit by deviating
unilaterally.  If an agent knew that the others were playing
according to an equilibrium profile (and would not change their
behavior), it would have no incentive to deviate.  Using the notation
we have outlined here, we can define a Nash equilibrium to be a
strategy profile $\sigma$ such that, for all $n \in N$ and all other
strategies $\sigma'_n \in \Sigma_n$, $G_n(\sigma_n,\sigma_{-n}) \geq
G_n(\sigma'_n,\sigma_{-n})$.

We can also define a notion of an approximate equilibrium, in which
each agent's incentive to deviate is small.  An
$\epsilon$\emph{-equilibrium} is a strategy profile $\sigma$ such that
no agent can improve its expected payoff by more than $\epsilon$ by
unilaterally deviating from $\sigma$.  In other words, for all $n \in
N$ and all other strategies $\sigma'_n \in \Sigma_n$,
$G_n(\sigma'_n,\sigma_{-n})- G_n(\sigma_n,\sigma_{-n}) \leq \epsilon$.
Unfortunately, finding an $\epsilon$-equilibrium is not necessarily a
step toward finding an exact equilibrium: the fact that $\sigma$ is an
$\epsilon$-equilibrium does not guarantee the existence of an exact
equilibrium in the neighborhood of $\sigma$.

\subsection{Normal-Form Games}

A normal-form game defines a simultaneous-move multi-agent scenario.
Each agent independently selects an action and then receives a payoff
that depends on the actions selected by all of the agents.  More
precisely, let $G$ be a normal-form game with a set $N$ of agents.
Each agent $n\in N$ has a discrete action set $A_n$ and a payoff array
$G_n$ with entries for every action profile in $A=\prod_{n\in N} A_n$ ---
that is, for joint actions $\ba=(a_{n_1},a_{n_2},\ldots,a_{n_{|N|}})$ of all
agents.  We use $A_{-n}$ to refer to the joint actions of agents in $N \setminus\{n\}$.

\subsubsection{Strategy Representation}
If agents are restricted to choosing actions deterministically, an
equilibrium is not guaranteed to exist.  
If, however, agents are allowed to independently randomize over actions, then
the seminal result of game theory~\cite{Nas51} guarantees the existence
of a \emph{mixed strategy} equilibrium.  A mixed strategy $\sigma_n$
is a probability distribution over $A_n$.

The strategy set $\Sigma_n$ is therefore defined to be the
probability simplex of all mixed strategies.  The \emph{support} of a mixed
strategy is the set of actions in $A_n$ that have non-zero
probability.  A strategy $\sigma_n$ for agent $n$ is said to be a
\emph{pure strategy} if it has only a single action in its support ---
pure strategies correspond exactly to the deterministic actions in
$A_n$.  The set $\Sigma$ of mixed strategy profiles is $\prod_{n \in
N} \Sigma_n$, a product of simplices.  A mixed strategy for a single
agent can be represented as a vector of probabilities, one for each
action.  For notational simplicity later on, we can concatenate all these
vectors and regard a mixed strategy profile $\sigma \in \Sigma$ as
a single $m$-vector, where $m=\sum_{n\in N}|A_n|$.  The vector is
indexed by actions in $\cup_{n\in N} A_n$, so for an action $a\in
A_n$, $\sigma_a$ is the probability that agent $n$ plays action $a$.
(Note that, for notational convenience, every action is associated with a particular agent;
different agents cannot take the ``same'' action.)  

\subsubsection{Payoffs}
A mixed strategy profile induces a joint distribution over action
profiles, and we can compute an expectation of payoffs with respect to
this distribution.  We let $G_n(\sigma)$
represent the expected payoff to agent $n$ when all agents behave
according to the strategy profile $\sigma$.  We can calculate this value by
\begin{equation}\label{eq:normal-form-payoff}
G_n(\sigma) = \sum_{\ba \in A} G_n(\ba) \prod_{k \in N} \sigma_{a_k}\,\,.
\end{equation}
In the most general case (a fully mixed strategy profile, in which every ), this sum
includes every entry in the game array $G_n$, which is exponentially
large in the number of agents.  

\subsection{Extensive-Form Games}

An extensive-form game is represented by a tree.  
The game proceeds sequentially from the
root.  Each non-leaf node in the tree
corresponds to a choice either of an agent or of nature; outgoing
branches represent possible actions to be taken at the node.  For each of 
nature's choice nodes, the game definition includes a probability distribution
over the outgoing branches (these are points in the game at which something happens
randomly in the world at large).  Each leaf $z \in Z$ of the tree is an outcome, and is
associated with a vector of payoffs $G(z)$, where $G_n(z)$ denotes the
payoff to agent $n$ at leaf $z$.  The choices of the agents and of nature
dictate which path of the tree is followed.

The choice nodes belonging to each agent are partitioned into
information sets; each information set is a set of states among
which the agent cannot distinguish.  Thus, an agent's strategy must
dictate the same behavior at all nodes in the same information set.  The
set of agent $n$'s information sets is denoted $I_n$, and the set of
actions available at information set $i\in I_n$ is denoted $A(i)$.  We
define an \emph{agent history} $H_n(\node)$ for a node $\node$ in the
tree and an agent $n$ to be a sequence containing pairs $(i,a)$ of the
information sets belonging to $n$ traversed in the path from the root to $\node$ 
(excluding the information set in which $\node$ itself is contained), and
the action selected by $n$ at each one.  Since actions are unique to information sets (the ``same'' action can't be taken at two different information sets), we can also omit the information sets and represent a history as an ordered tuple of actions only.  Two nodes have the same
agent-$n$ history if the paths used to reach them are
indistinguishable to $n$, although the paths may differ in other ways, such
as nature's decisions or the decisions of other agents.  We make the
common assumption of \emph{perfect recall}: an agent does not forget
information known nor choices made at its previous decisions.  More
precisely, if two nodes $\node,\node'$ are in the same information set
for agent $n$, then $H_n(\node) = H_n(\node')$.

\begin{figure}
\begin{center}
\includegraphics[height=3.0in]{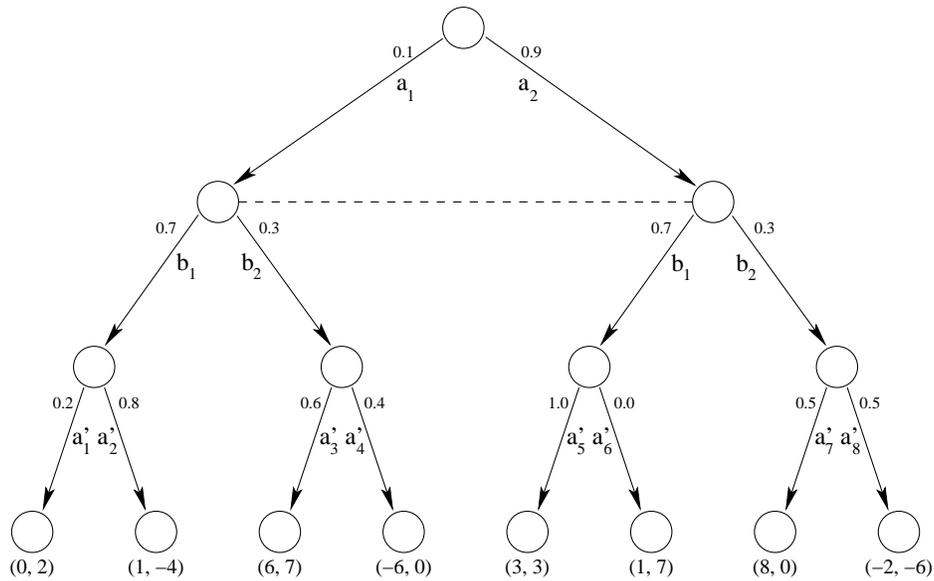}
\end{center}
\caption{A simple 2-agent extensive-form game.}
\label{fig:gametree}
\end{figure}

\begin{example}
\label{ex:gametree}
In the game tree shown in \figref{fig:gametree}, there are two agents,
Alice and Bob.  Alice first chooses between actions $a_1$ and $a_2$,
Bob next chooses $b_1$ or $b_2$, and then Alice chooses between two
of the set $\{a_1', a_2', \dots, a_8'\}$ (which pair depends on 
Bob's choice).  Information sets are indicated by nodes connected with dashed
lines.  Bob is unaware of Alice's actions, so both of his nodes are in
the same information set.  Alice is aware at the bottom level of both
her initial action and Bob's action, so each of her nodes is in a
distinct information set.  Edges have been labeled with the
probability that the agent whose action it is will follow it; note
that actions taken at nodes in the same information set must have the
same probability distribution associated with them.  There are eight
possible outcomes of the game,
each labeled with a pair of payoffs to Alice and Bob, respectively.  
\end{example}

\subsubsection{Strategy Representation}
\label{sec:enablingstrat}
Unlike the case of normal-form games, there are several quite
different choices of strategy representation for extensive-form games.  
One convenient formulation is in terms of
\emph{behavior strategies}.  A \emph{behavior profile} $b$ assigns to
each information set $i$ a distribution over the actions $a\in A(i)$.
The probability that agent $n$ takes action $a$ at information set
$i\in I_n$ is then written $b(a|i)$.  If $\node$ is a node in $i$,
then we can also write $b(a|\node)$ as an abbreviation for $b(a|i)$.

Our methods primarily employ a variant
of the \emph{sequence form} representation \cite{KolMeg92,vSt96,Rom62}, which
is built upon the behavior strategy representation.  In sequence form, 
a strategy $\sigma_n$ for an agent $n$ is represented as a \emph{realization plan}, 
a vector of real values.  Each value, or \emph{realization probability}, 
in the realization plan corresponds to a distinct history (or \emph{sequence}) $H_n(\node)$ that agent $n$ has, 
over all nodes $\node$ in the game tree.  Some of these sequences may only be 
partial records of $n$'s behavior in the game --- proper prefixes of larger sequences.  
The strategy representation employed by \GW (and by ourselves) is equivalent to the sequence form 
representation restricted to 
\emph{terminal sequences}: those which are agent-$n$ histories of at least one leaf node. 
We shall henceforth refer to this modified strategy representation simply as 
``sequence form,'' for the sake of simplicity.  

For agent $n$, then, we consider a realization plan 
$\sigma_n$ to be a vector of the realization probabilities of terminal sequences.  For an outcome 
$z$, $\sigma(H_n(z))$, abbreviated $\sigma_n(z)$, is the probability 
that agent $n$'s choices allow the realization of outcome $z$ --- 
in other words, the product of agent $n$'s behavior probabilities along the history
$H_n(z)$, $\prod_{(i,a)\in H_n(z)}b(a|i)$.  Several different outcomes 
may be associated with the same terminal sequence,
so that agent $n$ may have fewer realization probabilities than there 
are leaves in the tree.  The set of
realization plans for agent $n$ is therefore a subset of
$\reals^{\ell_n}$, where $\ell_n$, the number of distinct terminal sequences for
agent $n$, is at most the number of leaves in the tree.  

\begin{example}
In the example above, Alice has eight terminal sequences, one for
each of $a_1', a_2', \dots, a_8'$ from her four information
sets at the bottom level.  The history for one such last action is
$(a_1,a_3')$.  The realization probability $\sigma(a_1,a_3')$
is equal to $b(a_1)b(a_3'|a_1,b_2) = 0.1 \cdot 0.6 = 0.06$.  Bob has
only two last actions, whose realization probabilities are exactly his
behavior probabilities.
\end{example}

When all realization probabilities are non-zero,
realization plans and behavior strategies are in one-to-one 
correspondence.
(When some probabilities are zero, many possible behavior strategy
profiles might correspond to the same realization plan, as described by \citeR{KolMeg92};
this does not affect the work presented here.) From a behavior strategy
profile $b$, we can easily calculate
the realization probability $\sigma_n(z)=\prod_{(i,a)\in H_n(z)}b(a|i)$.
To understand the reverse transformation, note that we can also map 
behavior strategies to full realization plans defined on non-terminal
sequences (as they were originally defined by \citeR{KolMeg92}) by defining
$\sigma_n(h)=\prod_{(i,a)\in h}b(a|i)$; intuitively, 
$\sigma_n(h)$ is the probability 
that agent $n$'s choices allow the realization of partial sequence $h$.  
Using this observation, we can compute a behavior strategy from
an extended realization plan: if (partial) sequence
$(h,a)$ extends sequence $h$ by one action, namely action $a$ at information
set $i$ belonging to agent $n$, then we can compute $b(a|i)=\frac{\sigma_n(h,a)}
{\sigma_n(h)}$.  The extended realization probabilities can be computed
from the terminal realization probabilities by a recursive procedure
starting at the leaves of the tree and working upward: at information set
$i$ with agent-$n$ history $h$ (determined uniquely by perfect recall),  
$\sigma_n(h) = \sum_{a \in A(i)} \sigma_n(h,a)$.  

As several different information sets can have the same agent-$n$
history $h$, $\sigma_n(h)$ can be computed in multiple ways.  In order
for a (terminal) realization plan to be valid, it must satisfy the
constraint that all choices of information sets with agent-$n$ history
$h$ must give rise to the same value of $\sigma_n(h)$.  More formally,
for each partial sequence $h$, we have the constraints that for all
pairs of information sets $i_1$ and $i_2$ with $H_n(i_1)=H_n(i_2)=h$,
$\sum_{a \in A(i_1)} \sigma_n(h,a) = \sum_{a \in A(i_2)} \sigma_n(h,a)$.
In the game tree of \exref{ex:gametree}, consider Alice's realization
probability $\sigma_A(a_1)$.  It can be expressed as either
$\sigma_A(a_1,a_1')+\sigma_A(a_1,a_2') = 0.1\cdot 0.2+0.1\cdot 0.8$
or $\sigma_A(a_1,a_3')+\sigma_A(a_1,a_4')=0.1\cdot 0.6+0.1\cdot 0.4$,
so these two sums must be the same.

By recursively defining each realization probability as a sum of
realization probabilities for longer sequences, all constraints can be
expressed in terms of terminal realization probabilities;
in fact, the constraints are linear in these probabilities.  There are
several further constraints: all probabilities must be nonnegative, and,
for each agent $n$, $\sigma_n(\emptyset)=1$, where $\emptyset$ (the empty
sequence) is the agent-$n$ history of the first information set that agent
$n$ encounters.  This latter constraint simply enforces that
probabilities sum to one.  Together, these linear constraints define a
convex polytope $\Sigma$ of legal terminal realization plans.  

\subsubsection{Payoffs}
If all agents play according to $\sigma \in \Sigma$, the payoff to
agent $n$ in an extensive-form game is
\begin{equation}\label{eq:sequence-form-payoff}
G_n(\sigma) = \sum_{z \in Z} G_n(z) \prod_{k \in N} \sigma_{k}(z)\,\,,
\end{equation}
where here we have augmented $N$ to include nature for notational
convenience.  This is simply an expected sum of the payoffs over all
leaves.  For each agent $k$, $\sigma_k(z)$ is the product of the
probabilities controlled by $n$ along the path to $z$; thus,
$\prod_{k \in N} \sigma_{k}(z)$ is the multiplication of all
probabilities along the path to $z$, which is precisely the
probability of $z$ occurring.  
Importantly, this expression has a similar multi-linear form to the
payoff in a normal-form game, using realization plans rather
than mixed strategies.

Extensive-form games can be expressed
(inefficiently) as normal-form games, so they too are guaranteed to
have an equilibrium in mixed strategies.  In an extensive-form game
satisfying perfect recall, any mixed strategy profile can be
represented by a payoff-equivalent behavior profile, and hence
by a realization plan \cite{Kuh53}.  

\section{Structured Game Representations}
\label{sec:structrep}
The artificial intelligence community has recently introduced structured representations
that exploit independence relations 
in games in order to represent them compactly.
Our methods address 
two of these representations: graphical games \cite{KeaLitSin01}, a structured class of 
normal-form games, and MAIDs \cite{KolMil01}, a structured class of extensive-form games.  

\subsection{Graphical Games}
The size of the payoff arrays required to describe a normal-form game
grows exponentially with the number of agents.  In order to avoid
this blow-up, \citeA{KeaLitSin01}
introduced the framework of \emph{graphical games}, a more structured
representation inspired by probabilistic graphical models.  Graphical
games capture local structure in multi-agent interactions, allowing a
compact representation for scenarios in which each agent's payoff is only
affected by a small subset of other agents.  Examples of interactions
where this structure occurs include agents that interact along
organization hierarchies and agents that interact according to
geographic proximity.

A graphical game is similar in definition to a normal-form game, but
the representation is augmented by the inclusion of an interaction
graph with a node for each agent.  The original definition assumed an
undirected graph, but easily generalizes to directed graphs.  An edge
from agent $n'$ to agent $n$ in the graph indicates that agent $n$'s
payoffs depend on the action of agent $n'$.  More precisely, we define
$\family_n$ to be the set of agents consisting of $n$ itself and its
parents in the graph.  Agent $n$'s payoff function $G_n$ is an array indexed
only by the actions of the agents in $\family_n$.  Thus, the
description of the game is exponential in the in-degree of the graph
and not in the total number of agents.  In this case, we use
$\Sigma^f_{-n}$ and $A^f_{-n}$ to refer to strategy profiles and
action profiles, respectively, of the agents in $\family_n \setminus \{n\}$.

\begin{example}
\label{ex:roadgg}
Suppose $2L$ landowners along a road running north to south are
deciding whether to build a factory, a residential neighborhood, or a
shopping mall on their plots.  The plots are laid out along the road
in a $2$-by-$L$ grid; half of the agents are on the east side
($e_1,\ldots,e_L$) and half are on the west side ($w_1,\ldots,w_L$).
Each agent's payoff depends only on what it builds and what its
neighbors to the north, south, and across the road build.  For example,
no agent wants to build a residential neighborhood next to a factory.
Each agent's payoff matrix is indexed by the actions of at most
four agents (fewer at the ends of the road) and has $3^4$ entries, as
opposed to the full $3^{2L}$ entries required in the equivalent normal
form game.  (This example is due to \citeR{VicKol02}.)
\end{example}

\subsection{Multi-Agent Influence Diagrams}
The description length of extensive-form games can also grow exponentially 
with the number of agents.  In many situations, this large tree can be
represented more compactly.  \emph{Multi-agent influence diagrams}
(MAIDs)~\cite{KolMil01} allow a structured representation of games
involving time and information by extending \emph{influence
  diagrams}~\cite{HowMat84} to the multi-agent case.  

MAIDs and influence diagrams derive much of their syntax and semantics
from the Bayesian network framework.  A MAID compactly
represents a certain type of extensive-form game in much the same
way that a Bayesian network compactly represents a joint probability
distribution.  For a thorough treatment of Bayesian networks, we refer
the reader to the reference by \citeA{Cowetal99}.

\subsubsection{MAID Representation} 
Like a Bayesian network, a MAID defines a directed acyclic graph whose nodes correspond to random variables.  These random variables are partitioned into sets: a set $\sX$ of \emph{chance variables} whose values are chosen by nature, represented in the graph by ovals; for each agent $n$, a set $\sD_n$ of \emph{decision variables} whose values are chosen by agent $n$, represented by rectangles; and for each agent $n$, a set $\sU_n$ of \emph{utility variables}, represented by diamonds.  Chance and decision variables have, as their domains, finite sets of possible actions.  We refer to the domain of a random variable $V$ by $\dom(V)$. 
For each chance or decision variable $V$, the graph defines a parent set $\parents_V$ of those variables on whose values the choice at $V$ can depend.  Utility variables have finite sets of real payoff values for their domains, and are not permitted to have children in the graph; they represent components of an agent's payoffs, and not game state.  

The game definition supplies each chance variable $X$ with a conditional probability distribution (CPD) $P(X|\parents_X)$, conditioned on the values of the parent variables of $X$.  The semantics for a chance variable are identical to the semantics of a random variable in a Bayesian network; the CPD specifies the probability that an action in $\dom(X)$ will be selected by nature, given the actions taken at $X$'s parents.  The game definition also supplies a utility function for each utility node $U$.  The utility function maps each instantiation $\parass \in \dom(\parents_U)$ deterministically to a real value $U(\parass)$.  For notational and algorithmic convenience, we can regard this utility function as a CPD $P(U|\parents_U)$ in which, for each $\parass \in \dom(\parents_U)$, the value $U(\parass)$ has probability $1$ in $P(U|\parass)$ and all other values have probability $0$ (the domain of $U$ is simply the finite set of possible utility values).  At the end of the game, agent $n$'s total payoff is the sum of the utility received from each $U_n^i\in\sU_n$ (here $i$ is an index variable).  Note that each component $U_n^i$ of agent $n$'s payoff depends only on a subset of the variables in the MAID; the idea is to compactly decompose $a$'s payoff into additive pieces.  

\subsubsection{Strategy Representation}
The counterpart of a CPD for a decision node is a \emph{decision rule}.  A decision rule for a decision variable $D_n^i \in \sD_n$ is a function, specified by $n$, mapping each instantiation $\parass \in \dom(\parents_{D_n^i})$ to a probability distribution over the possible actions in $\dom(D_n^i)$.  A decision rule is identical in form to a conditional
probability distribution, and we can refer to it 
using the notation $P(D_n^i|\parents_{D_n^i})$.  As with the semantics for a chance node, the decision rule specifies the probability that agent $n$ will take any particular action in $\dom(D_n^i)$, having seen the actions taken at $D_n^i$'s parents.  
An assignment of decision rules to all $D_n^i \in \sD_n$ comprises a strategy for agent $n$.  
Once agent $n$ chooses a strategy, $n$'s behavior at $D_n^i$ depends only on the actions taken at $D_n^i$'s parents.  $\parents_{D_n^i}$ can therefore be regarded as the set of nodes whose values are visible to $n$ when it makes its choice at $D_n^i$. Agent $n$'s choice of \emph{strategy} may well take other nodes into account; but during actual game play, all nodes except those in $\parents_{D_n^i}$ are invisible to $n$. 

%
\begin{figure}
\begin{center}
\begin{tabular}{ccc}
\includegraphics[width=1.2in]{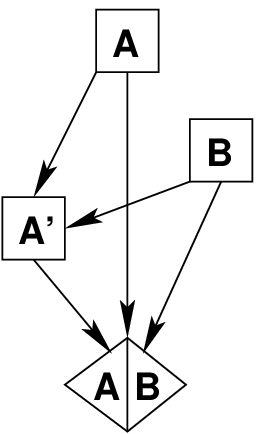} &
\hspace{0.5in} &
\includegraphics[width=3.0in]{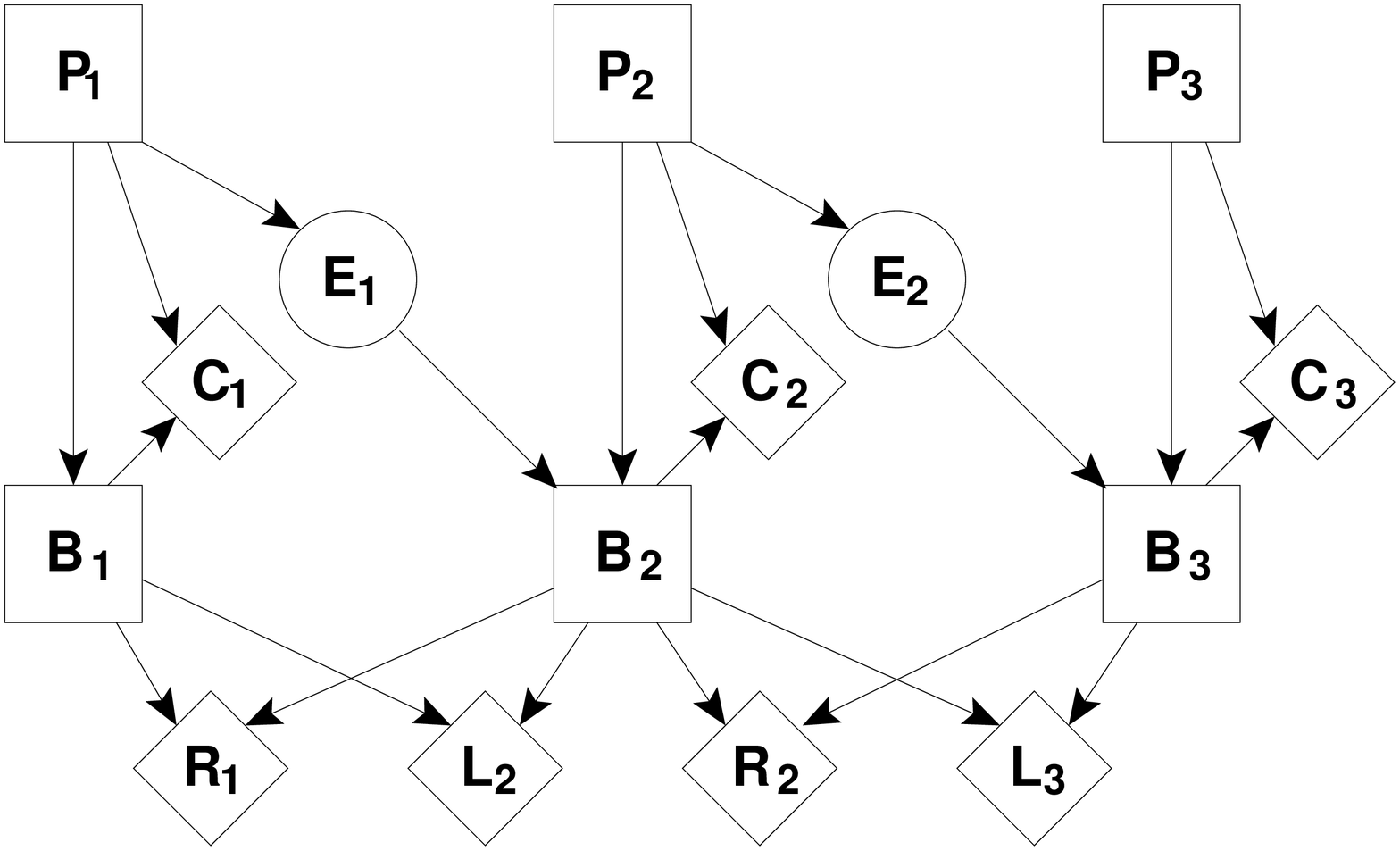} \\
(a) & & (b)  
\end{tabular}
\end{center}
\caption{
(a)
A simple MAID equivalent to the extensive form game in \figref{fig:gametree}.
(b)
A two-stage road game with three agents.
}
\label{fig:simplemaid}
\label{fig:twostageroad}
\end{figure}

\begin{example}
The extensive-form game considered in \exref{ex:gametree} 
can be represented by
the MAID shown in 
\subfigref{fig:simplemaid}{a}.  Alice and
Bob each have an initial decision to make without any information
about previous actions; then Alice has another decision to make in
which she is aware of Bob's action and her own.  Alice and Bob each have only one
utility node (the two are condensed into a single node in the graph, for
the sake of brevity), whose payoff structure is wholly general
(dependent on every action in the game) and thus whose possible values are
exactly the values from the payoff vectors in the extensive-form game.
\end{example}

\begin{example}
\label{ex:twostageroad}
\subfigref{fig:twostageroad}{b} shows a more complicated MAID of a somewhat
more realistic scenario.  Here, three landowners along a road are
deciding whether to build a store or a house.  Their payoff depends
only on what happens adjacent to them along the road.  Their decision
proceeds in two stages: the planning stage and the building stage.
The second landowner, for instance, has the two decision variables $P_2$
and $B_2$.  He receives a certain penalty from the utility node $C_2$ if he
builds the opposite of what he had planned to build.  But after planning,
he learns something about what his neighbor to the left has planned.
The chance node $E_1$ represents noisy espionage; it transmits the action
taken at $P_1$.  After learning the value of $E_1$, it may be in the second
landowner's interests to deviate from his plan, even if it means
incurring the penalty.  It is in his interest to start a trend that
distinguishes him from previous builders but which subsequent builders
will follow: the utility node $L_2$ rewards him for building the opposite
of what was built at $B_1$, and the utility node $R_2$ rewards him if the
third landowner builds the same thing he does at $B_3$.

Note that this MAID exhibits perfect recall, because the choice made at a 
planning stage is visible to the agent when it makes its next choice at the 
building stage.  
\end{example}

\subsubsection{Payoffs}
Under a particular strategy profile $\sigma$ --- that is, a tuple of strategies for all players --- all decision nodes have CPDs specified.  Since chance and utility nodes are endowed with CPDs already, the MAID therefore induces a fully-specified Bayesian network $\BN_\sigma$ with variables $\sV=\sX \cup \sD \cup \sU$ and the same directed graph as the MAID.  By the chain rule for Bayesian networks, $\BN_\sigma$ induces a joint probability distribution $P_{\sigma}$ over all the variables in $\sV$ by $P_{\sigma}(\sV)=\prod_{V \in \sV} P(V| \parents_V)$, with  CPDs for chance and utility variables given by the MAID definition and CPDs for decision variables given by $\sigma$.  For a game $G$ represented as a MAID, the expected payoff that agent $n$ receives under $\sigma$ is the expectation of $n$'s utility node values with respect to this distribution: 
\begin{align*}
G_n(\sigma) &= \sum_{U_n^i \in \sU_n} E_{P_{\sigma}}[U_n^i] \\
            &= \sum_{U_n^i \in \sU_n} \sum_{u\in \dom(U_n^i)} u\cdot P_\sigma(u) \text{.}
\end{align*}
We show in Section~\ref{sec:exstruct} that this 
and other related
expectations can be calculated efficiently using Bayesian network inference algorithms, giving a substantial performance increase over the
calculation of payoffs in the extensive-form game.

\subsubsection{Extensive Form Strategy Representations in MAIDs}
%
A MAID provides a compact definition of an extensive-form game.  We
note that, although this correspondence between MAIDs and extensive
form games provides some intuition about MAIDs, the details of the
mapping are not relevant to the remainder of the discussion.  We 
therefore briefly review this construction, referring to the work of \citeA{KolMil01} for details.  

The game tree associated with a MAID is a full, balanced tree, with
each path corresponding to a complete assignment of the chance and
decision nodes in the network.  Each node in the tree corresponds
either to a chance node or to a decision node of one of the
players, with an outgoing branch for each possible action at that node.  
All nodes at the same depth in the tree correspond to the same MAID node. 
We assume that the nodes along a path in the tree are ordered consistently with the
ordering implied by the directed edges in the MAID, so that if a MAID node
$X$ is a parent of a MAID node $Y$, the tree branches on $X$ before it
branches on $Y$.  The information sets for tree nodes
associated with a decision node $D_n^i$ correspond to assignments to the
parents $\parents_{D_n^i}$: all tree nodes corresponding to
$D_n^i$ with the same assignment to $\parents_{D_n^i}$ are
in a single information set.  We note that, by construction, the
assignment to $\parents_{D_n^i}$ was determined earlier in the tree,
and so the partition to information sets is well-defined.  For
example, the simple MAID in \subfigref{fig:simplemaid}{a} expands into
the much larger game tree that we saw earlier in
\figref{fig:gametree}.

Translating in the opposite direction, from extensive-form games to
MAIDs, is not always as natural.  If the game tree is unbalanced, then
we cannot simply reverse the above process.  However, with care, it is
possible to construct a MAID that is no larger than a given
extensive-form game, and that may be exponentially smaller in the number
of agents.  The details are fairly technical, and we omit them here in
the interest of brevity.

Despite the fact that a MAID will typically be much more compact than the equivalent extensive-form game, the strategy representations of the two turn out to be equivalent and of equal size.  A decision rule for a decision variable $D_n^i$ assigns a distribution over actions to each joint assignment to $\parents_{D_n^i}$, just as a behavior strategy assigns a distribution over actions to an information set in an extensive form game --- as discussed above, each assignment to the parents of $D_n^i$ is an information set.  A strategy
profile for a MAID --- a set of decision rules for every decision variable
--- is therefore
equivalent to a set of behavior strategies for every information
set, which is simply a behavior profile.

If we make the assumption of perfect recall, then, since MAID
strategies are simply behavior strategies, we can represent them in
sequence form.  Perfect recall requires that no agent forget anything
that it has learned over the course of the game.  In the MAID formalism, the
perfect recall assumption is equivalent to the following constraint: if
agent $n$ has two decision nodes $D_n^i$ and $D_n^j$, with the second
occurring after the first, then all parents of $D_n^i$ (the information $n$
is aware of in making decision $D_n^i$) and $D_n^i$ itself must be parents
of $D_n^j$.  This implies that agent $n$'s final decision node
$D_n^d$ has, as parents, all of $n$'s previous decision nodes and
their parents. Then a joint assignment to $D_n^d \cup \parents_{D_n^d}$ precisely
determines agent $n$'s sequence of information sets and actions
leading to an outcome of the game --- the agent-$n$ history of the
outcome.  

%
The realization probability for a particular sequence is computed by multiplying all behavior
strategy probabilities for actions in that sequence.  In MAIDs,
a sequence corresponds to a joint assignment to $D_n^d \cup \parents_{D_n^d}$, and the behavior strategy probabilities for this sequence are entries consistent with this assignment in the decision rules for agent $n$.  
We can therefore derive all of agent $n$'s realization probabilities at once by
multiplying together, as conditional probability distributions,
the decision rules of each of agent $n$'s
decision nodes in the sequence --- when multiplying conditional probability distributions, only those entries whose assignments are consistent with each other are multiplied.  
Conversely, given a realization 
plan, we can derive the behavior strategies and hence the
decision rules according to the method outlined for extensive-form
games.

In the simple MAID example in \subfigref{fig:simplemaid}{a}, the terminal sequences
are the same as in the equivalent extensive-form game.
In the road example in \subfigref{fig:twostageroad}{b}, agent $2$ has $8$
terminal sequences; one for each joint assignment to his
final decision node ($B_2$) and its parents ($E_1$ and $P_2$).
Their associated realization probabilities are given by multiplying the decision rules at
$P_2$ and at $B_2$.


\section{Computational Complexity}
\label{sec:cplex}
When developing algorithms to compute equilibria efficiently, the question
naturally arises of how well 
one can expect these algorithms to perform.
The complexity of computing Nash equilibria has been studied for some time.  
\citeA{GilZem89}
first showed that it is NP-hard to find more than one Nash equilibrium
in a normal-form game, and \citeA{ConSan03} recently utilized a simpler
reduction to arrive at this result and several others in the same
vein.  Other recent hardness results pertain to restricted subclasses
of normal-form games \cite<e.g.,>{ChuHal01,CodSte05}.
However, these results apply only to 2-agent normal-form games.  While it is true
that proving a certain subclass of a class of problems to be NP-hard also
proves the entire class to be NP-hard (because NP-hardness is a measure of
\emph{worst}-case complexity), such a proof might tell us very little about the
complexity of problems outside the subclass.  
This issue is particularly apparent in the problem of computing equilibria, 
because games can grow along two distinct axes: the number of agents,
and the number of actions per agent.  The hardness results of \citeA{ConSan03} 
apply only as the number of actions per agent increases.  
Because 2-agent
normal-form games are (fully connected) graphical games, 
these results apply to graphical games. 
 
However, we are more interested in the hardness of graphical games as
the number of agents increases, rather than the number of actions per
agent.  It is graphical games with large numbers of agents that
capture the most structure --- these are the games for which the graphical
game representation was designed.  
In order to prove results about the
asymptotic hardness of
computing equilibria along this more interesting (in this setting) axis of
representation size, we require a different reduction.  
Our proof, like a number of previous hardness proofs for games \cite<e.g.,>{ChuHal01,ConSan03,CodSte05}, reduces 3SAT to equilibrium computation.  However, in these previous proofs, variables in 3SAT instances are mapped to \emph{actions} (or sets of actions) in a game with only $2$ players, whereas in our reduction they are mapped to \emph{agents}. 
Although differing in approach, our reduction is very much in the
spirit of the reduction appearing in the work of \citeA{ConSan03}, and many of the corollaries of
their main result also follow from ours (in a form adapted to graphical games).  

\begin{theorem}
\label{thmnphard}
For any constant $d \geq 5$, and $k \geq 2$, the problem of deciding whether a
graphical game with a family size at most $d$ and at most $k$ actions per
player has more than one Nash equilibrium is NP-hard.
\end{theorem}
\begin{proof}
Deferred to $\appref{nphardproof}$.
\end{proof}

In our reduction, all games that have more than one equilibria have at least one
pure strategy equilibrium.  This immediately gives us

\begin{corollary}
\label{cor:gridnp}
It is NP-hard to determine whether a graphical game has more than one
Nash equilibrium in discretized strategies with even the coarsest
possible granularity.
\end{corollary}

Finally, because graphical games can be represented as (trivial) MAIDs, 
in which each agent has only a single parentless decision node and a single utility node,
and each agent's utility node has, as parents, the decision nodes of the graphical game family
of that agent, we obtain the following corollary.

\begin{corollary}
\label{cor:maidnp}
It is NP-hard to determine whether a MAID with constant family size at least 6 has
more than one Nash equilibrium.
\end{corollary}

\section{Continuation Methods}
\label{sec:contmeth}
Continuation methods form the basis of our algorithms for solving each
of these structured game representations.  We begin with a high-level
overview of continuation methods, referring the reader to the work of~\citeA{Wat00}
for a more detailed discussion.  

Continuation methods work by solving
a simpler perturbed problem and then tracing the solution as the
magnitude of the perturbation decreases, converging to a solution for
the original problem.
More precisely, let $\lambda$ be a scalar parameterizing a continuum
of perturbed problems.  When $\lambda=0$, the perturbed problem is the
original one; when $\lambda=1$, the perturbed problem is one for which
the solution is known.  Let $\bw$ represent the vector of real values
of the solution.  For any perturbed problem defined by $\lambda$, we
characterize solutions by the equation $F(\bw,\lambda)=\bzero$, where
$F$ is a real-valued vector function of the same dimension as $\bw$
 (so that $\bzero$ is a vector of
zeros).  The function $F$ is such that $\bw$ is a solution to the problem 
perturbed by $\lambda$ if and only if $F(\bw,\lambda)=\bzero$.

The continuation method traces solutions along the level set of
solution pairs $(\bw,\lambda)$ satisfying $F(\bw,\lambda)=\bzero$.
Specifically, if we have a solution pair $(\bw, \lambda)$, we would
like to trace that solution to a nearby solution.  Differential
changes to $\bw$ and $\lambda$ must cancel out so that $F$ remains
equal to $\bzero$.  

If $(\bw,\lambda)$ changes in the direction of a unit vector $\bu$, then
$F$ will change in the direction $\nabla F \cdot \bu$, where $\nabla F$
is the Jacobian of $F$ (which can also be written
$\begin{bmatrix}\nabla_\bw F & \nabla_\lambda F\end{bmatrix}$).  We
want to find a direction $\bu$ such that $F$ remains unchanged, i.e., equal
to $\bzero$.  Thus, we need to solve the matrix equation
\begin{equation}\label{eq:continuation-path}
\begin{bmatrix}\nabla_\bw F & \nabla_\lambda F\end{bmatrix}
\begin{bmatrix}d\bw \\ d\lambda\end{bmatrix} = 0\,\,.
\end{equation}
Equivalently, changes $d\bw$ and $d\lambda$ along the path must obey
$\nabla_\bw F\cdot d\bw = -\nabla_\lambda\,F \cdot d\lambda$.  
Rather than inverting the matrix
$\nabla_\bw F$ in solving this equation, we use the adjoint
$\adjoint(\nabla_\bw F)$, which is still defined when $\nabla_\bw F$
has a null space of rank $1$.  The adjoint is the matrix of cofactors:
the element at $(i,j)$ is $(-1)^{i+j}$ times the determinant of the
sub-matrix in which row $i$ and column $j$ have been removed.  When
the inverse is defined, $\adjoint (\nabla_\bw F) = \det(\nabla_\bw F)
[\nabla_\bw F]^{-1}$.  In practice, we therefore set $d\bw =
-\adjoint(\nabla_\bw F) \cdot \nabla_\lambda F$ and
$d\lambda=\det(\nabla_\bw F)$.  
If the Jacobian
$[\nabla_\bw F \ \ \nabla_\lambda F]$ has a null-space of rank 1
everywhere, the curve is uniquely defined.

The function $F$ should be constructed so that the curve starting at $\lambda=1$ is
guaranteed to cross $\lambda=0$, at which point the corresponding
value of $\bw$ is a solution to the original problem.  A continuation
method begins at the known solution for $\lambda=1$ .  The null-space
of the Jacobian $\nabla F$ at a current solution $(\bw,\lambda)$
defines a direction, along which the solution is moved by a small
amount.  The Jacobian is then recalculated and the process repeats,
tracing the curve until $\lambda=0$.  The cost of each step in this
computation
is at least 
cubic in the size of $\bw$,
due to the required matrix operations.
However, the Jacobian itself may in general be much more difficult
to compute.  
\citeA{Wat00} provides some simple examples of continuation methods.

\section{Continuation Methods for Games}
\label{sec:contgnm}
We now review the work of \GW on applying the continuation method to
the task of finding equilibria in games.  They provide continuation
methods for both normal-form and extensive-form games.  These
algorithms form the basis for our extension to structured games,
described in the next section.  The continuation methods perturb the
game by giving agents fixed bonuses, scaled by $\lambda$, for each of
their actions, independently of whatever else happens in the game.  If
the bonuses are large enough (and unique), they dominate the original
game structure, and the agents need not consider their opponents'
actions.  There is thus a unique pure-strategy equilibrium easily
determined by the bonuses at $\lambda=1$.  The continuation method can
then be used to follow a path in the space of $\lambda$ and
equilibrium profiles for the resulting perturbed game, decreasing
$\lambda$ until it is zero; at this point, the corresponding strategy
profile is an equilibrium of the original game.

\subsection{Continuation Method for Normal-Form Games}
We now make this intuition more precise, beginning with normal-form games.

\subsubsection{Perturbations}
A perturbation vector $\bb$ is a vector of $m$ values chosen at random, one for each
action in the game.  The bonus $b_a$ is given to the agent $n$
owning action $a$ for playing $a$, independently of whatever else
happens in the game.  Applying this perturbation to a target game $G$
gives us a new game, which we denote $G\oplus\bb$, in which, for each
$a\in A_n$, and for any $\bt\in A_{-n}$, $(G\oplus\bb)_n(a,\bt) =
G_n(a,\bt)+b_a$.  If $\bb$ is made sufficiently large, then
$G\oplus\bb$ has a unique equilibrium, in which each agent plays the
pure strategy $a$ for which $b_a$ is maximal.

\subsubsection{Characterization of Equilibria}
\label{sec:eqchar}
In order to apply \eqref{eq:continuation-path}, we need to
characterize the equilibria of perturbed games as the zeros of a
function $F$.
Using a structure theorem of
\citeA{KohMer86},
\GW show
that the continuation method path deriving from their
equilibrium characterization leads to convergence for all perturbation
vectors except those in a set of measure zero.
We present only the equilibrium characterization here; proofs of the
characterization and of the method's convergence are given by \citeA{GovWil03}.

We first define an auxiliary vector function $V^G(\sigma)$, indexed by
actions, of the payoffs to each agent for deviating from $\sigma$ to
play a single action.  
We call $V^G$ the \emph{deviation function}.
The element $V^G_a(\sigma)$ corresponding to a
single action $a$, owned by an agent $n$, is the payoff to agent $n$
when it deviates from the mixed strategy profile $\sigma$ by playing the
pure strategy for action $a$:
\begin{equation}\label{eq:Vnormal-form}
V^G_a(\sigma)=
  \sum_{\bt\in  A_{-n}}G_n(a,\bt)\!\!\! \prod_{k\in N\setminus\{n\}}\!\!\!\sigma_{t_k}.
\end{equation}
It can also be viewed as the component of agent $n$'s payoff that it
derives from action $a$, under the strategy profile $\sigma$.  Since
bonuses are given to actions independently of $\sigma$, the effect of
bonuses on $V^G$ is independent of $\sigma$.  $V^G_a$ measures the
payoff for deviating and playing $a$, and bonuses are given for
precisely this deviation, so $V^{G\oplus \bb}(\sigma) =
V^G(\sigma)+\bb$.
 
We also utilize the
\emph{retraction operator} $R:\reals^m \to \Sigma$
defined by \citeA{GulPeaSta93}, which
maps an arbitrary $m$-vector $\bw$ to the point in
the space $\Sigma$ of mixed strategies which is nearest to $\bw$ in
Euclidean distance.  Given this operator, the equilibrium
characterization is as follows.  
\begin{lemma}\cite{GulPeaSta93}
\label{lem:eqchar}
If $\sigma$ is a strategy profile of $G$, then
$\sigma=R(V^G(\sigma)+\sigma)$ iff $\sigma$ is an equilibrium.
\end{lemma}
Although we omit the proof, we will give some intuition for why this result is true.
Suppose $\sigma$
is a fully-mixed equilibrium; that is, every action has non-zero
probability.  For a single agent $n$, $V^G_a(\sigma)$ must be the
same for all actions $a\in A_n$, because $n$ should not have any
incentive to deviate and play a single one of them.  Let $V_n$ be the
vector of entries in $V^G(\sigma)$ corresponding to actions of $n$,
and let $\sigma_n$ be defined similarly. $V_n$ is a scalar multiple of
$\bone$, the all-ones vector, and the simplex $\Sigma_n$ of $n$'s mixed
strategies is defined by $\bone^T\bx = 1$, so $V_n$ is orthogonal to
$\Sigma_n$.  $V^G(\sigma)$ is therefore orthogonal to $\Sigma$, so
retracting $\sigma+V^G(\sigma)$ onto $\Sigma$ gives precisely
$\sigma$.  
In the reverse direction, if $\sigma$ is a fully-mixed strategy
profile satisfying $\sigma=R(V^G(\sigma)+\sigma)$, then $V^G(\sigma)$ must be
orthogonal to the polytope of mixed strategies.  Then, for each agent,
every pure strategy has the same payoff.  Therefore, $\sigma$ is in fact
an equilibrium.  A little more care must be taken when dealing with actions
not in the support.  We refer to \citeA{GulPeaSta93} for the details.  

According to \lemref{lem:eqchar}, we can define an equilibrium as a
solution to the equation $\sigma = R(\sigma + V^G(\sigma))$.  
On the other hand, if 
$\sigma = R(\bw)$ for some $\bw \in \reals^m$, 
we have the equivalent condition that $\bw=R(\bw)+V^G(R(\bw))$; $\sigma$ is 
an equilibrium iff this condition is satisfied, as can easily be verified.
We can therefore
search for a point $\bw\in\reals^m$ which satisfies this equality, in
which case $R(\bw)$ is guaranteed to be an equilibrium. 

The form of our continuation equation is then
\begin{equation}
F(\bw,\lambda)=\bw-R(\bw)-\left(V^G\left(R\left(\bw\right)\right)+\lambda \bb\right)\,\,.
\end{equation}
We have that $V^G+\lambda \bb$ is the deviation function for the perturbed
game $G\oplus\lambda \bb$, so $F(\bw,\lambda)$ is zero if and only
if $R(\bw)$ is an equilibrium of $G\oplus\lambda \bb$.  At $\lambda=0$ the
game is unperturbed, so $F(\bw,0)=\bzero$ iff $R(\bw)$ is an
equilibrium of $G$.

\subsubsection{Computation}
The expensive step in the continuation method is the calculation of the
Jacobian $\nabla_\bw F$, required for the computation that maintains the
constraint of \eqref{eq:continuation-path}.  Here, we have that $\nabla_\bw F
= I-(I+\nabla V^G) \nabla R$, where $I$ is the $m \times m$ identity matrix.
The hard part is the calculation of $\nabla V^G$.  For pure strategies
$a \in A_n$ and $a' \in A_{n'}$, for $n' \neq n$, the value at location $(a,a')$ in
$\nabla V^G(\sigma)$ is equal to the expected payoff to agent $n$ when it
plays the pure strategy $a$, agent $n'$ plays the pure strategy $a'$, and
all other agents act according to the strategy profile $\sigma$:
\begin{align}
\nabla V^G_{a,a'}(\sigma) &=
\frac{\partial}{\partial \sigma_{a'}}\sum_{\bt\in A_{-n}}G_{n}(a,\bt)\!\!\!
     \prod_{k\in N\setminus\{n\}}\!\!\!\!\sigma_{\bt_k} \notag \\
 & = \sum_{\bt\in A_{-n,n'}}\!\!G_{n}(a,a',\bt)\!\!\!\!
     \prod_{k\in N\setminus\{n,n'\}}\!\!\!\!\!\sigma_{\bt_k}\,\,. \label{eq:gradVnf}
\end{align}
If both $a \in A_n$ and $a' \in A_n$, $\nabla V^G_{a,a'}(\sigma) = 0$.  

Computing \eqref{eq:gradVnf} requires a large number of
multiplications; the sum is over
the space $A_{-n,n'} = \prod_{k\in
N\setminus\{n,n'\}}A_i$, which is exponentially large in the number of agents.

\subsection{Continuation Method for Extensive-Form Games}
\label{sec:extensive}
The same method applies to extensive-form games, using the sequence form
strategy representation.  
\subsubsection{Perturbations}
As with normal-form games, the game is perturbed by the bonus vector
$\bb$.  Agent $n$ owning sequence $h$ is paid an additional bonus
$\bb_h$ for playing $h$, 
independently of whatever else
happens in the game.  Applying this perturbation gives us a new game
$G\oplus\bb$ in which, for each $z \in Z$, $(G\oplus\bb)_n(z) =
G_n(z)+\bb_{H_n(z)}$.

If the bonuses are large enough and unique, \GW show that once again
the perturbed game has a unique pure-strategy equilibrium (one in
which all realization probabilities are $0$ or $1$).  However,
calculating it is not as simple as in the case of normal-form games.
Behavior strategies must be calculated from the leaves upward by a
recursive procedure, in which at each step the agent who owns the node
in question chooses the action that results in the sequence with the
largest bonus.  Since all actions below it have been recursively
determined, each action at the node in question determines an outcome.  
The realization plans can be derived from this
behavior profile by the method outlined in \secref{sec:enablingstrat}.  

\subsubsection{Characterization of Equilibria}
Once more, we first define a vector function capturing the 
benefit of deviating
from a given strategy profile, indexed by sequences:
\begin{equation}\label{eq:Vef}
V^G_h(\sigma)=
 \sum_{z\in Z_h}
     G_{n}(z)\!\!\!\!\prod_{k\in N\setminus\{n\}}\!\!\!\! \sigma_k(z),
\end{equation}
where $Z_h$ is the set of leaves that are consistent with the
sequence $h$.  
The interpretation of $V^G$ is not as natural as in
the case of normal-form games, as it is not possible for an agent
to play one sequence to the exclusion of all others; its possible actions
will be partially determined by the actions of other agents.  In this case,
$V^G_h(\sigma)$ can be regarded as the \emph{portion} of its payoff that agent
$n$ receives for playing sequence $h$, unscaled by agent $n$'s own probability of
playing that sequence.  
As with normal-form games, the vector of bonuses is
added directly to $V^G$, so $V^{G\oplus\bb}=V^G+\bb$.

The retraction operator $R$ for realization plans is defined in
the same way as for normal-form strategies: it takes a general vector
and projects it onto the nearest point in the valid region of realization plans.  
The constraints defining this space are
linear, as discussed in \secref{sec:enablingstrat} .  We can therefore express them as a
constraint matrix $C$ with $C \sigma = 0$ for all valid profiles
$\sigma$.  In addition, all probabilities must be greater than or
equal to zero.  To calculate $\bw$, we must find a $\sigma$
minimizing $(w-\sigma)^T(w-\sigma)$, the (squared) Euclidean distance
between $\bw$ and $\sigma$, subject to $C \sigma = \bzero$ and $\sigma
\geq \bzero$.  This is a quadratic program (QP), which can be solved
efficiently using standard methods.  The Jacobian of the retraction is
easily computable from the set of active constraints.

The equilibrium characterization for realization plans is
now surprisingly similar to that of mixed strategies in normal-form games;
\GW show that, as before, equilibria are characterized by $\sigma =
R(\sigma + V^G(\sigma))$, where now $R$ is the retraction for sequence form
and $V^G$ is the deviation function.  The continuation equation $F$ takes
exactly the same form as well.  

\subsubsection{Computation}
The key property of the reduced sequence-form strategy representation is that
the deviation function is a multi-linear function of the extensive-form
parameters, as shown in \eqref{eq:Vef}.  The elements of the Jacobian
$\nabla V^G$ thus also have the same general structure.  In particular,
the element corresponding to sequence $h$ for agent $n$ and
sequence $h'$ for agent $n'$ is
\begin{align}
\nabla V^G_{h,h'}(\sigma) & =
 \frac{\partial}{\partial \sigma_{h'}} \sum_{z\in Z_h}
     G_n(z)\!\!\!\prod_{k\in N\setminus\{n\}}\!\!\!\sigma_k(z)\notag \\
& = 
 \sum_{z\in Z_{h,h'}}
     G_n(z)\!\!\!\!\prod_{k\in N\setminus\{n,n'\}}\!\!\!\!\! \sigma_k(z) \label{eq:gradVext}
\end{align}
where $Z_{h,h'}$ is the set of leaves that are consistent with
the sequences $h$ (for agent $n$) and $h'$ (for agent $n'$).   
$Z_{h,h'}$ is the empty set (and hence $\nabla V^G = 0$) if $h$ and
$h'$ are incompatible.  \eqref{eq:gradVext} is precisely analogous to
\eqref{eq:gradVnf} for normal-form games.  We have a sum over outcomes of
the utility of the outcome multiplied by the strategy probabilities for all
other agents.  Note that this sum is over the leaves of the tree, which may
be exponentially numerous in the number of agents.  

One additional subtlety,
which must be addressed by any method for equilibrium computation in extensive-form games,
relates to zero-probability actions.  Such
actions induce a probability of zero for entire trajectories in the
tree, possibly leading to equilibria based on unrealizable threats.
Additionally, for information sets that occur with zero probability,
agents can behave arbitrarily without disturbing the equilibrium
criterion, resulting in a continuum of equilibria and a possible
bifurcation in the continuation path.  This prevents our methods from
converging.  We therefore constrain all realization probabilities to be
greater than or equal to $\epsilon$ for some small $\epsilon>0$.  This
is, in fact, a requirement for \GW's equilibrium characterization to
hold.  The algorithm thus looks for an $\epsilon$-\emph{perfect
equilibrium}~\cite{FudTir91}: a strategy profile $\sigma$ in which each
component is constrained by $\sigma_s \geq \epsilon$, and each agent's
strategy is a best response among those satisfying the constraint.
Note that this is entirely different from an $\epsilon$-equilibrium.
An $\epsilon$-perfect equilibrium always exists, as long as $\epsilon$
is not so large as to make the set of legal strategies empty.  
An $\epsilon$-perfect equilibrium can be
interpreted as an equilibrium in a perturbed game in which agents have a
small probability of choosing an unintended action.  A limit of
$\epsilon$-perfect equilibria as $\epsilon$ approaches $0$ is a
\emph{perfect equilibrium}~\cite{FudTir91}: a refinement of the basic
notion of a Nash equilibrium.  As $\epsilon$ approaches $0$, the
equilibria found by \GW's algorithm therefore converge to an exact
perfect equilibrium, by continuity of the variation in the
continuation method path.  Then for $\epsilon$ small enough, there is
a perfect equilibrium in the vicinity of the found $\epsilon$-perfect
equilibrium, which can easily be found with local search.

\subsection{Path Properties}
\begin{figure}
\begin{center}
\includegraphics[width=3in]{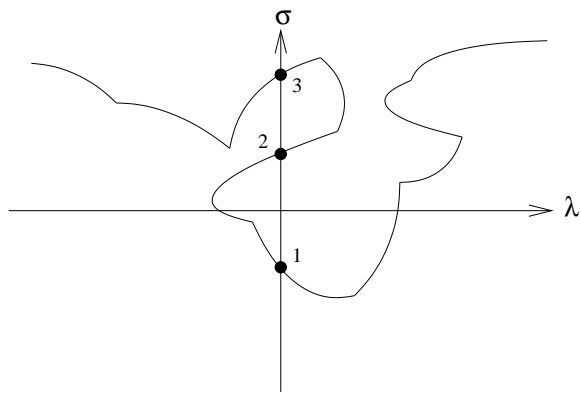}
\end{center}
\caption{ An abstract diagram of the path.  The horizontal axis
 represents $\lambda$ and the vertical axis represents the space of
 strategy profiles (actually multidimensional).  The algorithm starts
 on the right at $\lambda=1$ and follows the dynamical system until
 $\lambda=0$ at point 1, where it has found an equilibrium of the
 original game.  It can continue to trace the path and find the
 equilibria labeled 2 and 3.  }
\label{fig:path}
\end{figure}
In the case of normal-form games, \GW show, using the structure
theorem of \citeA{KohMer86}, that the path of the algorithm is a
one-manifold without boundary with probability one over all choices
for $\bb$.  They provide an analogous structure theorem that
guarantees the same property for extensive-form games.
\subfigref{fig:path}{a} shows an abstract representation of the path
followed by the continuation method.  \GW show that the path must 
cross the $\lambda=0$ hyperplane at least once,
yielding an equilibrium.  In fact, the path may cross multiple times,
yielding many equilibria in a single run.  As the path must eventually
continue to the $\lambda=-\infty$ side, it will find an odd number of
equilibria when run to completion.

In both normal-form and extensive-form games, the path is piece-wise
polynomial, with each piece corresponding to a different support set
of the strategy profile.  These pieces are called \emph{support
cells}.  The path is not smooth at cell boundaries due to
discontinuities in the Jacobian of the retraction operator, and hence
in $\nabla_\bw F$, when the support changes.  Care must be taken
to step up to these boundaries exactly when following the path; at
this point, the Jacobian for the new support can be calculated and the
path can be traced into the new support cell.

In the case of two agents, the path is piece-wise linear and, rather
than taking steps, the algorithm can jump from corner to corner along
the path.  
When this algorithm is applied to a two-agent game and a
particular bonus vector is used (in which only a single entry is non-zero), 
the steps from support cell to
support cell that the algorithm takes are 
identical to the pivots
of the Lemke-Howson algorithm~\cite{LemHow64} for two-agent general-sum
games, and the two algorithms find precisely the same set of solutions~\cite{GovWil02}.
Thus, the continuation method is a strict generalization of the Lemke-Howson
algorithm that allows different perturbation rays and games of more
than two agents.  

This process is described in more detail in the pseudo-code for the
algorithm, presented in \figref{fig:pseudo}.

\subsection{Computational Issues}
\label{sec:compissues}
Guarantees of convergence apply only as long as we stay on the path
defined by the dynamical system of the continuation method.  However,
for computational purposes, discrete steps must be taken.  As a
result, error inevitably accumulates as the path is traced, so that $F$
becomes slightly non-zero.  
\GW use several simple techniques to combat
this problem.  We adopt their techniques, and introduce one of our own:
we employ an adaptive step size, taking smaller steps when error accumulates 
quickly and larger ones when it
does not.  When $F$ is nearly linear (as it is, for example, when very few
actions are in the support of the current strategy profile), this technique speeds 
computation significantly.  

\GW use two different techniques to remove
error once it has accumulated.  Suppose we are at a point
$(w,\lambda)$ and we wish to minimize the magnitude of $F(w,\lambda)=
w-V^G(R(w))+\lambda \bb+R(w)$.  There are two values we might change:
$w$, or $\lambda \bb$.  We can change the first without affecting the
guarantee of convergence, so every few steps we run a local Newton
method search for a $w$ minimizing 
$|F(w,\lambda)|$.  If this search
does not decrease error sufficiently, then we perform what \GW call a
``wobble'': we change the perturbation vector (``wobble'' the
continuation path) to make the current solution consistent.  If we set
$\bb=[w-V^G(R(w))-R(w)]/\lambda$, the equilibrium characterization
equation is immediately satisfied.  Changing the perturbation vector
invalidates any theoretical guarantees of convergence.  However, it is
nonetheless an attractive option because it immediately reduces error
to zero. Both the local Newton method and the ``wobbles'' are described
in more detail by \citeA{GovWil03}.

These techniques can potentially send the algorithm into a cycle, and
in practice they occasionally do.  However, they are necessary for
keeping the algorithm on the path.  If the algorithm cycles, random
restarts and a decrease in step size can improve convergence.  More
sophisticated path-following algorithms might also be used, and in
general could improve the success rate and execution time of the
algorithm.

\subsection{Iterated Polymatrix Approximation}
\label{sec:ipa}
Because perturbed games may themselves have a large number of
equilibria, and the path may wind back and forth through any number of
them, the continuation algorithm can take a while to trace its way
back to a solution to the original game.  We can speed up the
algorithm using an initialization procedure based on the
\emph{iterated polymatrix approximation} (\ipa) algorithm of \GW.  A
\emph{polymatrix game} is a normal-form game in which the payoffs to an
agent $n$ are equal to the sum of the payoffs from a set of two-agent
games, each involving $n$ and another agent.  Because polymatrix games
are a linear combination of two-agent normal-form games, they reduce
to a linear complementarity problem and can be solved quickly using
the Lemke-Howson algorithm \cite{LemHow64}.

For each agent $n\in N$ in a polymatrix game, the payoff array is a
matrix $B^n$ indexed by the actions of agent $n$ and of each other
agent; for actions $a\in A_n$ and $a'\in A_{n'}$, $B^n_{a,a'}$ is the
payoff $n$ receives for playing $a$ in its game with agent $n'$, when
$n'$ plays $a'$.  
Agent $n$'s total payoff is the sum of the payoffs it receives from its
games with each other agent, $\sum_{n' \neq n} \sum_{a \in A_n, a' \in A_{n'}} \sigma_a \sigma_{a'}B^n_{a,a'}$.  
Given a normal-form game $G$ and a strategy profile
$\sigma$, we can construct the polymatrix game $P_\sigma$ whose payoff
function has the same Jacobian at $\sigma$ as $G$'s
by setting
\begin{equation}
B^n_{a,a'}=\nabla V^G_{a,a'}(\sigma)\,\,.
\end{equation}

The game $P_\sigma$ is a linearization of $G$ around
$\sigma$: its Jacobian is the same everywhere.  \GW show that
$\sigma$ is an equilibrium of $G$ if and only if it is an equilibrium
of $P_\sigma$.  This follows from the equation $V^G(\sigma)=\nabla
V^G(\sigma)\cdot \sigma /(|N|-1)$, which holds for all $\sigma$.  To see
why it holds, consider the single element indexed by $a \in A_n$: 
\begin{align}
(\nabla V^G(\sigma)\cdot \sigma)_a &= \sum_{n' \in N\setminus\{n\}}\ \sum_{a' \in
     A_{n'}}\sigma_{a'}\!\!\!\sum_{\bt\in A_{-n,-n'}}\!\!\!G_n(a,a',\bt)\!\!\!\!
     \prod_{k\in N\setminus\{n,n'\}}\!\!\!\!\!\sigma_{\bt_k} \notag \\
&=  \sum_{n' \in N\setminus\{n\}} \!\ \ 
     \sum_{\bt\in A_{-n}}G_n(a,\bt)\!\!\!
     \prod_{k\in N\setminus\{n\}}\!\!\sigma_{\bt_k} \notag \\
&= (|N|-1)V^G(\sigma)_a. \notag
\end{align}
The equilibrium characterization equation can therefore be written
\begin{equation}
\notag \sigma=R\left(\sigma+\nabla V^G\left(\sigma\right)\cdot\sigma \left(|N|-1\right)\right)\text{.}
\end{equation}
$G$ and $P_\sigma$ have the same value
of $\nabla V$ at $\sigma$, and thus the same equilibrium
characterization function.  Then $\sigma$ satisfies one if and only if
it satisfies the other.

We define the mapping $p: \Sigma \to \Sigma$ such that $p(\sigma)$ is
an equilibrium for $P_\sigma$ (specifically, the first equilibrium found by
the Lemke-Howson algorithm).  If $p(\sigma) = \sigma$, then $\sigma$ is an
equilibrium of $G$.  The \ipa procedure of \citeA{GovWil04} aims to find
such a fixed point.  It begins with a randomly chosen strategy profile
$\sigma$, and then calculates $p(\sigma)$ by running the Lemke-Howson
algorithm; it adjusts $\sigma$ toward $p(\sigma)$ using an approximate
derivative estimate of $p$ built up over the past two iterations.  If
$\sigma$ and $p(\sigma)$ are sufficiently close, it terminates with an
approximate equilibrium.

\ipa is not guaranteed to converge.  However, in practice, it quickly
moves ``near'' a good solution.  It is possible at this point to
calculate a perturbed game close to the original game (essentially,
one that differs from it by the same amount that $G$'s polymatrix
approximation differs from $G$) for which the found approximate
equilibrium is in fact an exact equilibrium.  The continuation method
can then be run from this starting point to find an exact equilibrium
of the original game.
The continuation method is not guaranteed to converge from this starting 
point.  However, in practice we have always found it to converge, as long as IPA is 
configured to search for high quality equilibrium approximations.  Although 
there are no theoretical results on the required quality, IPA can
refine the starting point further if the continuation method fails.  
Our results show that 
the IPA quick-start substantially reduces the overall running time of our
algorithm.

We can in fact use any other approximate algorithm as a quick-start for
ours, also without any guarantees of convergence.  Given an
approximate equilibrium $\sigma$, the inverse image of $\sigma$ under
$R$ is defined by a set of linear constraints.  If we let
$w:=V^G(\sigma)+\sigma$, then we can use standard QP methods to retract
$w$ to the nearest point $w'$ satisfying these constraints, and let
$\bb:=w'-w$.  Then $\sigma=R(w')=R(V^G(\sigma)+\sigma+\bb)$, so we are on
a continuation method path.  Alternatively, we can choose $\bb$
by ``wobbling'', in which case we set $\bb:=[w-V^G(R(w))-R(w)]/\lambda$.  

\begin{figure}
\footnotesize
For an input game $G$: 
\begin{enumerate}
\item
Set $\lambda=1$, choose initial $\bb$ and $\sigma$ either by a
quick-start procedure ({\it e.g.},~\ipa) or by randomizing.  Set $\bw=V^G(\sigma)+\lambda
\bb+\sigma$.

\item While $\lambda$ is greater than some (negative) threshold ({\it
i.e.},~there is still a good chance of picking up another equilibrium):
\begin{enumerate}
\item
\label{code:newsupp}
Initialize for the current support cell: set the $\steps$ counter to
the number of steps we will take in crossing the cell, depending on
the current amount of error.  If $F$ is linear or nearly linear (if,
for example, the strategy profile is nearly pure, or there are only
$2$ agents), set $\steps=1$ so we will cross the entire cell.

\item
While $\steps\geq 1$:

\begin{enumerate}

\item 
\label{code:jac}
Compute $\nabla V^G(\sigma)$.

\item
Set $\nabla_\bw F(\bw,\lambda)=I-(\nabla V^G(\sigma)+I)\nabla R(\bw)$
(we already know $\nabla_\lambda F = -\bb$).  Set $d\bw =
\adjoint(\nabla_\bw F) \cdot \bb$ and $d\lambda=det(\nabla_\bw F)$.
These satisfy \eqref{eq:continuation-path}.

\item
Set $\delta$ equal to the distance we'd have to go in the direction of
$d\bw$ to reach the next support boundary.  We will scale $d\bw$ and
$d\lambda$ by $\delta / \steps$.

\item
If $\lambda$ will change signs in the course of the step, record an
equilibrium at the point where it is $0$.

\item
Set $\bw:=\bw+ d\bw (\delta/ \steps)$ and $\lambda:=\lambda+ d\lambda (\delta /
\steps)$.

\item
If sufficient error has accumulated, use the local Newton method to
find a $\bw$ minimizing $|F(\bw,\lambda)|$.  If this does not reduce
error enough, increase $\steps$, thereby decreasing step size.  If we
have already increased $\steps$, perform a ``wobble'' and reassign $\bb$.

\item
Set $\steps := \steps - 1$.

\end{enumerate}
\end{enumerate}
\end{enumerate}
\caption{ Pseudo-code for the \cont algorithm. }
\label{fig:pseudo}
\end{figure}

\section{Exploiting Structure}
\label{sec:exstruct}
Our algorithm's continuation method foundation is the same for each
game representation, but the calculation of $\nabla V^G$ in
\coderef{code:jac} of the pseudo-code in \figref{fig:pseudo} is
different for each and consumes most of the time.  Both in normal-form
and (in the worst case) in extensive-form games, it requires
exponential time in the number of agents.  However, as we show in this
section, when using a structured representation such as a graphical
game or a MAID, we can effectively exploit the structure of the game
to drastically reduce the computational time required.

\subsection{Graphical Games}
Since a graphical game is also a normal-form game, the definition of
the deviation function $V^G$ in \eqref{eq:Vnormal-form} is the
same: $V^G_a(\sigma)$ is the payoff to agent $n$ for deviating from
$\sigma$ to play $a$ deterministically.  However, due to the structure
of the graphical game, the choice of strategy for an agent outside the
family of $n$ does not affect agent $n'$s payoff.  This observation allows us to
compute this payoff locally.  

\subsubsection{The Jacobian for Graphical Games}
We begin with the definition of
$V^G$ for normal-form games (modified slightly to account for the
local payoff arrays).  
Recall that
$A_{-n}^f$ is the set of action profiles of agents in $\family_n$ other
than $n$, and let $A_{-\family_n}$ be the set of action profiles of
agents not in $\family_n$.  Then we can divide a sum over full action profiles between
these two sets, switching from the normal-form version of $G^n$ to the
graphical game version of $G^n$, as follows:
\begin{align}
V^G_a(\sigma) & =  \sum_{\bt\in  A_{-n}}G_n(a,\bt)\!\!\! 
\prod_{k\in N\setminus\{n\}}\!\!\!\sigma_{\bt_k} \notag \\
& = \sum_{\bu\in  A^f_{-n}}G_n(a,\bu)\!\!\!\!\! 
\prod_{k\in \family_n\setminus\{n\}}\!\!\!\!\!\sigma_{\bu_k}\!\!\!
\sum_{\bv\in A_{-\family_n}} \prod_{j \in N\setminus\family_n}\!\!\!\!\sigma_{\bv_j}\,\,. \label{eq:Vgraph}
\end{align}
Note that the latter sum and product simply sum out a probability
distribution, and hence are always equal to $1$ due to the constraints
on $\sigma$.  They can thus be eliminated without changing the value
$V^G$ takes on valid strategy profiles.  However, their partial
derivatives with respect to strategies of agents not in $\family_n$
are non-zero, so they enter into the computation of $\nabla V^G$.

Suppose we wish to compute a row in the Jacobian matrix corresponding
to action $a$ of agent $n$.  We must compute the entries for each
action $a'$ of each agent $n'\in N$.  In the trivial case where $n'=n$
then $\nabla V^G_{a,a'}=0$, since $\sigma_a$ does not appear
anywhere in the expression for $V^G_a(\sigma)$.  We next compute
the entries for each action $a'$ of each other agent $n' \in
\family_{n}$.  In this case,
\begin{align}
\nabla V^G_{a,a'}(\sigma) & = \frac{\partial}{\partial \sigma_{a'}} \sum_{\bu\in  A^f_{-n}}G_n(a,\bu)\!\!\!\!\! 
\prod_{k\in \family_n\setminus\{n\}}\!\!\!\sigma_{\bu_k}\!\!
\sum_{\bv\in A_{-\family_n}} \prod_{j \in N\setminus\family_n}\!\!\!\!\sigma_{\bv_j} \label{eq:Vgder} \\
& = \sum_{\bu\in  A^f_{-n}}G_n(a,\bu)\frac{\partial}{\partial \sigma_{a'}} \!\! 
\prod_{k\in \family_n\setminus \{n\}}\!\!\!\sigma_{\bu_k} \cdot 1 \notag \\
& =   \sum_{\bt\in A^f_{-n,n'}}\!\!\!G_n(a,a',\bt)\!\!\!\!\!
     \prod_{k\in \family_n\setminus \{n,n'\}}\!\!\!\!\!\sigma_{\bt_k}
\text{, \ \ \ \ if $n'\in\family_n$} 
\,\,.
\label{eq:Vgfam}
\end{align}
We next compute the
entry for a single action $a'$ of an agent $n' \notin
\family_n$.  The derivative in \eqref{eq:Vgder} takes a different
form in this case; the variable in question is in the second
summation, not the first, so that we have
\begin{align}
\nabla V^G_{a,a'}(\sigma) & = \frac{\partial}{\partial \sigma_{a'}} \sum_{\bu\in  A^f_{-n}}G_n(a,\bu)\!\!\!\!\! 
\prod_{k\in \family_n\setminus \{n\}}\!\!\!\sigma_{\bu_k}\!\!
\sum_{\bv\in A_{-\family_n}} \prod_{j \in N\setminus\family_n}\!\!\!\!\sigma_{\bv_j} \notag \\
 & = \sum_{\bu\in  A^f_{-n}}G_n(a,\bu)\!\!\!\!\! 
\prod_{k\in \family_n\setminus \{n\}}\!\!\!\sigma_{\bu_k}\!\!
\sum_{\bv\in A_{-\family_n}} \frac{\partial}{\partial \sigma_{a'}} \prod_{j \in N\setminus\family_n}\!\!\!\!\sigma_{\bv_j} \notag \\
& = \sum_{\bu\in  A^f_{-n}}G_n(a,\bu) \!\! 
\prod_{k\in \family_n\setminus \{n\}}\!\!\!\sigma_{\bu_k} \cdot 1 
\text{, \ \ \ \ if $n'\not\in\family_n$}
\,\,.
\label{eq:gradVgg1}
\end{align}
Notice that this calculation does not depend on $a'$; therefore, it
is the same for each action of each other agent not in
$\family_n$.  We need not compute any more elements of the row.  
We can copy this value into all other columns of actions
belonging to agents not in $\family_n$.  

\subsubsection{Computational Complexity}
\label{sec:ggcplex}
Due to graphical game structure, the computation of $\nabla V^G(\sigma)$ 
takes time exponential only in the 
maximal family size of the game, and hence takes time polynomial in the
number of agents if the family size is 
constant.  In particular, our methods lead to the following theorem about the 
complexity of the continuation method for graphical games.

\begin{theorem}
\label{GGContComplexity}
The time complexity of computing the Jacobian of the deviation function
$\nabla V^G(\sigma)$ for a graphical game is
$O(f d^f |N|+ d^2|N|^2)$, where $f$ is the maximal family size 
and $d$ is the maximal number of actions per agent.  
\end{theorem}
\begin{proof}
Consider a single row in the Jacobian,
corresponding to a single action $a$ owned by a single agent $n$.
There are at most $d(f-1)$ entries in the row for actions owned
by other members of $\family_n$.  For one such action $a'$, the computation
of the Jacobian element $\nabla V^G_{a,a'}$ according to 
\eqref{eq:Vgfam} takes time $O(d^{f-2})$.  The total cost for all
such entries is therefore $O((f-1) d^{f-1})$.  There are then at most
$d(|N|-f)$ entries for actions owned by non-family-members.  The value
of $\nabla V^G_{a,a'}$ for each such $a'$ is the same.  It can be
calculated once in time $O(d^{f-1})$, then copied across the row in time
$d(|N|-f)$.  All in all, the computational cost for the row is
$O(f d^{f-1}+d |N|)$.  There are at most $d |N|$ rows, so the total
computational cost is $O(|N| f d^f + d^2|N|^2)$.
\end{proof}

Each iteration of the algorithm calculates $\nabla V^G(\sigma)$ once;
we have therefore proved that a single iteration takes time polynomial
in $|N|$ if $f$ is constant (in fact, matrix operations make the complexity
cubic in $|N|$).   However, as for normal-form games, there are 
no theoretical results about how many steps of the continuation method
are required for convergence.  

\subsection{MAIDs}
For graphical games, the exploitation of structure was straightforward.
We now turn to the more difficult problem of exploiting structure in
MAIDs.  We take advantage of two distinct sets of structural properties.  The
first, a coarse-grained structural measure known as \emph{strategic
relevance}~\cite{KolMil01}, has been used in previous computational
methods.  After decomposing a MAID according to strategic
relevance relations, we can exploit finer-grained structure by using
the extensive-form continuation method of \GW to solve each component's
equivalent extensive-form game.  In the next two sections, we describe
these two kinds of structure.  

\subsubsection{Strategic Relevance}
\begin{figure}
\begin{center}
\begin{tabular}{ccc}
\includegraphics[width=1.2in]{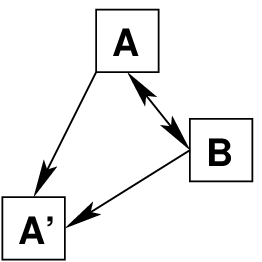} &
\hspace{0.5in} &
\includegraphics[width=3.0in]{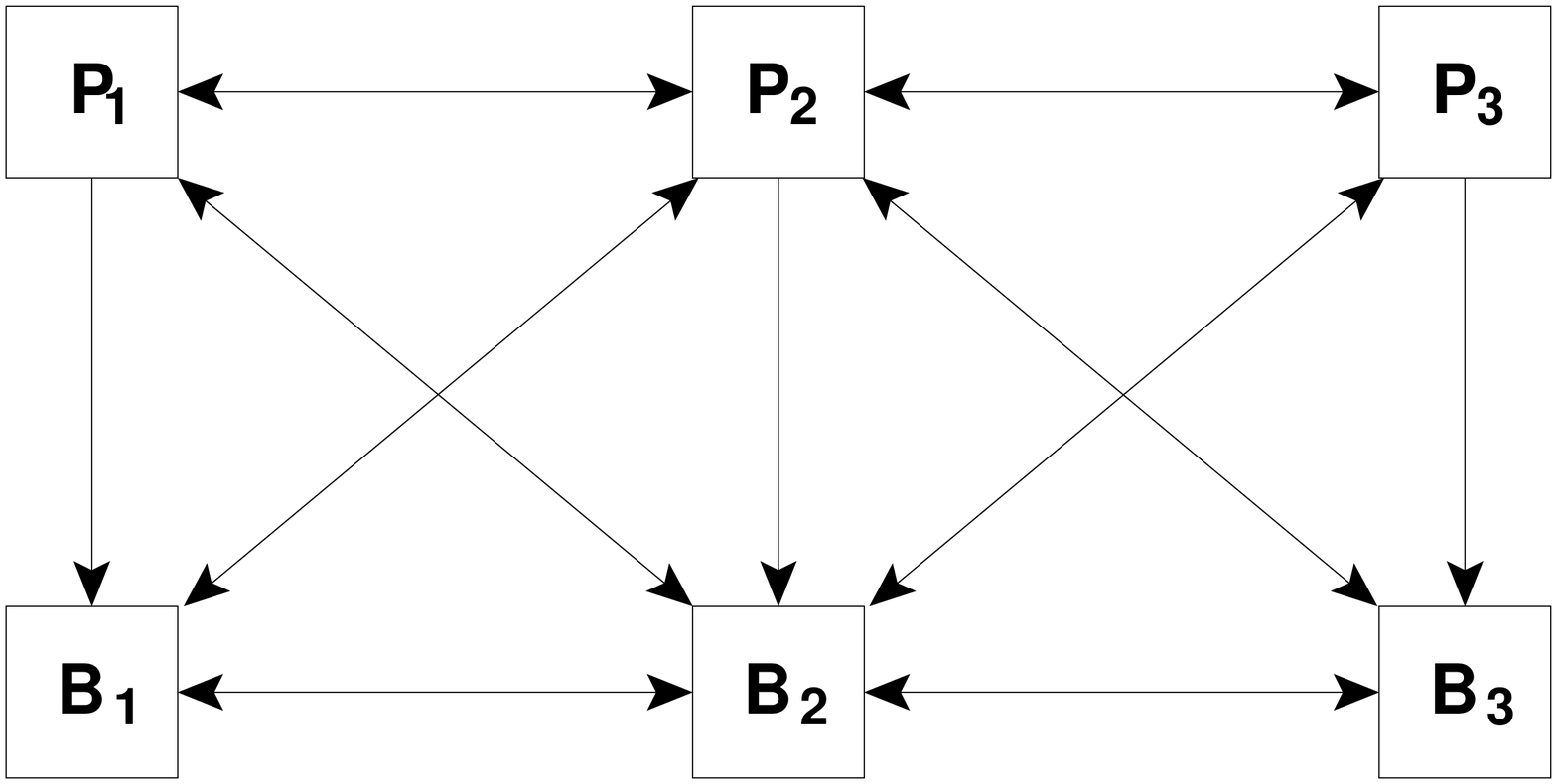} \\
(a) & & (b)  
\end{tabular}
\end{center}
\caption{The strategic relevance graphs for the MAIDs in
(a) \protect\subfigref{fig:simplemaid}{a}
and (b) \protect\subfigref{fig:twostageroad}{b}. }
\label{fig:relgraphs}
\end{figure}

Intuitively, a decision node $D_n^i$ is \emph{strategically relevant}
to another decision node $D_{n'}^j$ if agent $n'$, in order to optimize
its decision rule at $D_{n'}^j$, needs to know agent $n$'s decision rule
at $D_n^i$.  The relevance relation induces a directed graph known as
the \emph{relevance graph}, in which only decision nodes appear and an 
edge from node $D_{n'}^j$ to node $D_n^i$ is present iff $D_n^i$ is strategically relevant
to $D_{n'}^j$.  In the event that the relevance graph is acyclic, 
the decision rules can be optimized sequentially in any reverse topological order;
when all the children of a node $D_n^i$ have had their decision rules set,
the decision rule at $D_n^i$ can be optimized without regard for 
any other nodes.  

When cycles exist in the relevance graph, however, further steps must
be taken.  Within a \emph{strongly connected component (SCC)}, a set of nodes
for which a directed path between any two nodes exists in the relevance
graph, decision rules cannot be optimized sequentially---in any linear ordering
of the nodes in the SCC, some node must be optimized before one of its 
children, which is impossible. \citeA{KolMil01} show 
that a MAID can be decomposed into SCCs, which can then be solved
individually.  

For example, the relevance graph for the MAID in \subfigref{fig:simplemaid}{a}, shown in
\subfigref{fig:relgraphs}{a}, has one SCC consisting of $A$ and $B$, and
another consisting of $A'$.  In this MAID, we would first optimize the
decision rule at $A'$, as the optimal decision rule at $A'$ does not rely on
the decision rules at $A$ and $B$ --- when she makes her decision at $A'$,
Alice already knows the actions taken at $A$ and $B$, so she does not need to
know the decision rules that led to them.  Then we would turn $A'$ into a
chance node with CPD specified by the optimized decision rule and optimize the
decision rules at $A'$ and $B$.  The relevance graph for
\subfigref{fig:twostageroad}{b}, shown in \subfigref{fig:relgraphs}{b}, forms
a single strongly connected component.

The computational method of \citeA{KolMil01} stops at strategic
relevance: each SCC is converted into an equivalent extensive-form game and
solved using standard methods.  Our algorithm can be viewed as an augmentation
of their method: after a MAID has been decomposed into SCCs, we
can solve each of these SCCs using our methods, taking advantage of
finer-grained MAID structure within them to find equilibria
more efficiently.  The MAIDs on which we test our algorithms (including
the road MAID in \subfigrefn{fig:twostageroad}{b}) all have strongly
connected relevance graphs, so they cannot be decomposed (see
\subfigrefn{fig:relgraphs}{b} and \figref{fig:chaingame}).

\subsubsection{The Jacobian for MAIDs}
A MAID is equivalent to an extensive-form game, so its deviation
function $V^G$ is the same one defined in \eqref{eq:gradVext}.  Now, however, we can
compute the payoffs that make up the Jacobian $\nabla V^G$ more efficiently.  Consider
a payoff $G_n(z)$ to agent $n$ for outcome $z$.  The outcome $z$ is simply an assignment 
$\bx$ to all of the variables in the MAID.  The realization probability $\sigma_n(z)$
is the product of the probabilities for the decisions of agent $n$ in the assignment $\bx$,
so the product $\prod_{k \in N} \sigma_k(z)$ of all 
realization probabilities is simply the joint probability of the assignment.  The expected
payoff agent $n$ will receive under the strategy profile $\sigma$, 
$\sum_{z\in Z} G_n(z)\prod_{k \in N} \sigma_k(z)$, is therefore an expectation of $G_n(z)$. 
The expectation is with respect to the distribution $P_{\sigma}$ defined 
by the Bayesian network
$\BN_\sigma$ whose structure is the same as the MAID, with decision
node CPDs determined by $\sigma$.  

The entries of $\nabla V^G$ are not strictly expected payoffs, however.  \eqref{eq:gradVext}
can be rewritten as
\begin{equation}
\nabla V^G_{h,h'}(\sigma) = \sum_{z\in Z_{h,h'}}
     \frac{G_n(z) \prod_{k\in N}\sigma_k(z)}
          {\sigma_n(z)\sigma_{n'}(z)}\,\,.
\label{eq:gradVmaid}
\end{equation}
The expectation is of the quantity $G_n(z)/[\sigma_n(z)\sigma_{n'}(z)]$.
The payoff $G_n(z)$ is the sum of agent $n$'s utility nodes.  Due to
linearity of expectation, we can perform the computation separately for
each of agent $n$'s utility nodes, and then simply add up the separate
contributions.

We can therefore restrict our attention to computing the contribution of
a single utility node $U_n$ for each agent $n$.  
Furthermore, 
the value of $\sigma_n(z)$ depends only on the 
values of the set of nodes $\bD_n$ consisting of $n$'s decision nodes and
their parents.  Thus, instead of computing the probabilities for all
assignments to all variables, we need only compute the marginal joint
distribution over 
$U_n$,
 $\bD_n$, and $\bD_{n'}$.  From
this distribution, we can compute the contribution of $U_n$ to the
expectation in \eqref{eq:gradVmaid} for every pair of terminal sequences
belonging to agents $n$ and $n'$.  

\subsubsection{Using Bayesian Network Inference}
Our analysis above reduces the required computations significantly.
Rather than computing a separate expectation for every pair of
sequences $h,h'$, as might at first have seemed necessary,
we need only compute one marginal joint distribution over the variables 
in 
$\{U_n\}\cup \bD_n \cup \bD_{n'}$
 for every pair of agents $n,n'$. 
This marginal joint distribution is the one defined by the Bayesian
network $\BN_\sigma$.  Naively, this computation requires that we
execute Bayesian network inference $|N|^2$ times: once for each
ordered pair of agents $n,n'$.  In fact, we can exploit the
structure of the MAID to perform this computation much more
efficiently.  The basis for our method is the standard \emph{clique
tree algorithm} of \citeA{LauSpi88}.  
The clique tree algorithm is fairly complex, and a detailed
presentation is outside the scope of this paper.  We choose to treat
the algorithm as a black box, describing only those of its properties that are
relevant to understanding how it is used within our computation.  We
note that these details suffice to allow our method to be implemented
using one of the many off-the-shelf implementations of the clique tree
algorithm.  A reader wishing to understand the clique tree algorithm
or its derivation in more detail is referred to the reference by \citeA{Cowetal99} for
a complete description.

A clique tree for a Bayesian network
$\BN$ is a data structure defined over an undirected tree with a set
of nodes $\C$.  Each node $C_i \in \C$ corresponds to some subset of the variables 
in $\BN$, typically called a \emph{clique}.  The clique tree satisfies certain
important properties.  It must be \emph{family preserving}: for each
node $X$ in $\BN$, there exists a clique $C_i \in \C$ such that $(X\cup \parents_X) \subseteq C_i$.  
It also satisfies a
\emph{separation} requirement: if $C_2$ lies on the unique path from $C_1$ to
$C_3$, then, in the joint distribution defined by $\BN$, the
variables in $C_1$ must be conditionally independent of those in $C_3$
given those in $C_2$.  

\begin{figure}
\begin{center}
\begin{tabular}{m{3.1in}m{2.6in}}
\includegraphics[width=3.0in]{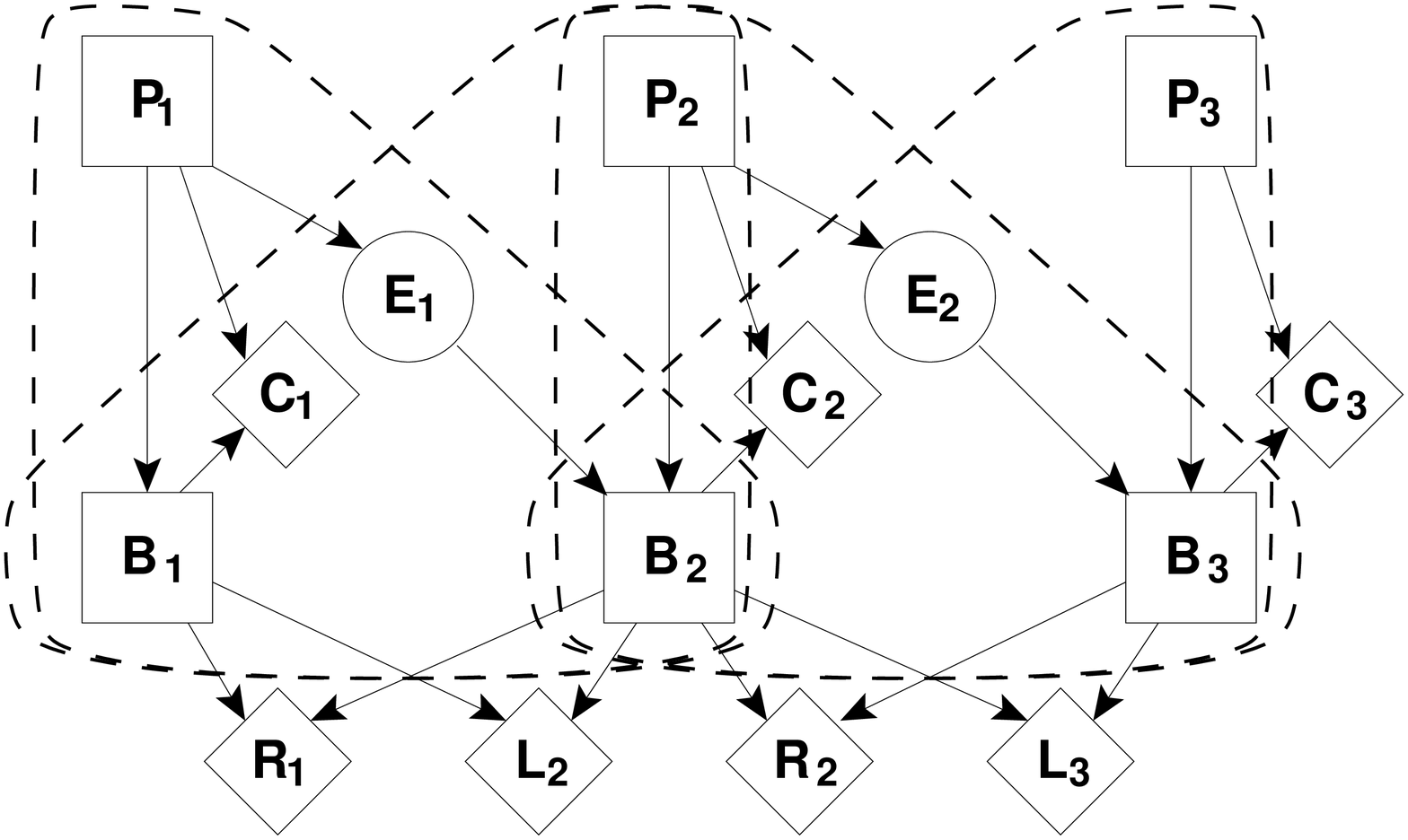} &
\includegraphics[width=2.5in]{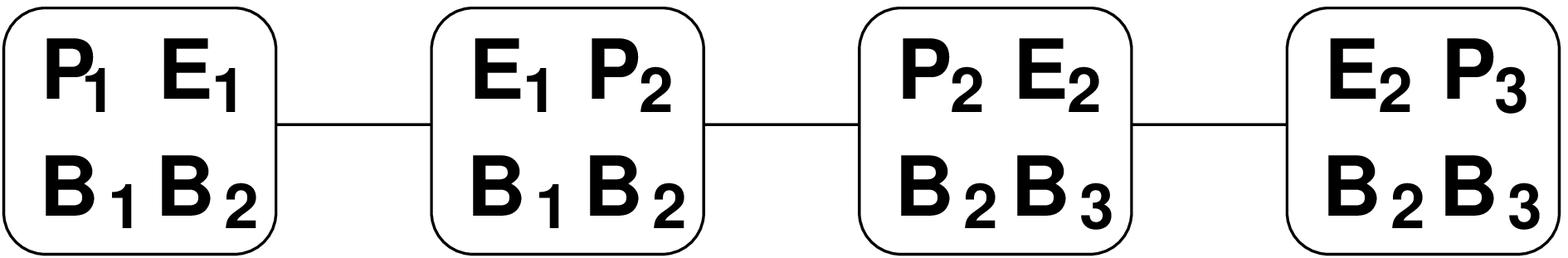} \\
\hspace{1.45in}(a) & \hspace{1.2in}(b)
\end{tabular}
\caption{ 
(a) A two-stage road MAID with three agents is
shown divided into cliques.  Each of the four cliques is surrounded by
a dashed line, and has three decision nodes and a chance node.
(b) The resultant clique tree.
}
\end{center}
\label{fig:cliquetree}
\end{figure}

The division of the 3-agent road MAID into cliques is shown in
\subfigref{fig:cliquetree}{a}.
This MAID has 4 cliques.  Notice that
every family is contained in a clique (including the families of
chance nodes and utility nodes).  The clique tree for this MAID is shown in
\subfigref{fig:cliquetree}{b}.

Each clique maintains a data structure called a \emph{potential}, a
table with an entry for each joint assignment to  
the variables in the clique.  
A table of this sort is more generally called a \emph{factor}.  Inference algorithms
typically use two basic operations on factors: factor product, and
factor marginalization.  If $\sF$ and $\sG$ are two
factors over the (possibly overlapping) sets of variables $\bX$ and
$\bY$, respectively, then we can define the product $\sF \sG$ to be a
new factor over $\bX \cup \bY$.  The entry in $\sF \sG$ for a
particular assignment to the variables in $\bX \cup \bY$ is the
product of the entries in $\sF$ and $\sG$ corresponding to the
restriction of the assignment to $\bX$ and $\bY$, respectively.  This
notion of multiplication corresponds to the way that conditional
probability distributions are multiplied.  We can also
\emph{marginalize}, or sum, a variable $X$ out of a factor $\sF$ over
$\bX$ in the same way in which we would sum a variable out of a joint
probability distribution.  The result is a factor $\sum_X \sF$  
over the variables in $\bX \backslash \{X\}$.  The entry for a particular
assignment to the variables in $\sum_X \sF$ is equal to the sum of all entries in $\sF$ 
compatible with that assignment --- one for each value of $X$.  

Because a factor has an entry for
every joint assignment to its variables, the size of the
potential for $C_i$ is exponential in $|C_i|$.  The clique
tree inference algorithm proceeds by passing messages, themselves factors,
from one clique
to another in the tree.  The messages are used to update the potential
in the receiving clique by factor multiplication.  After a process in which messages have been
sent in both directions over each edge in the tree, the tree is said
to be \emph{calibrated}; at this point, the potential of every clique
$C_i$ contains precisely the joint distribution over the variables in
$C_i$ according to $\BN$ 
(for details, we refer to the reference by \citeR{Cowetal99}).

We can use the clique tree algorithm to perform inference over
$\BN_\sigma$.  Consider the final decision node for agent $n$.
Due to the perfect recall assumption, all of $n$'s previous
decisions and all of their parents are also parents of this decision
node.  The family preservation property therefore implies that
$\bD_{n}$ is fully contained in some clique.  It also implies that
 the family of each utility node is contained in a clique.  The
expectation of \eqref{eq:gradVmaid} thus requires the computation of
the joint distribution over three cliques in the tree: the one
containing $\parents_{U_n}$, the one containing $\bD_n$, and the one
containing $\bD_{n'}$.  We need to compute this joint distribution
for every pair of agents $n,n'$.

The first key insight is that we can reduce this problem to one of
computing the joint marginal distribution for all pairs of cliques in the
tree.  Assume we have computed $P_{\BN}(C_i,C_j)$ for every pair of cliques
$C_i,C_j$.  Now, consider any triple of cliques $C_i,C_j,C_k$.  There
are two cases: either one of these cliques is on the path between the
other two, or not.  In the first case, assume without loss of generality
that $C_j$ is on the path from $C_i$ to $C_k$.  In this case, by the
separation requirement, we have that 
$
P_{\BN}(C_i,C_j,C_k) = P_{\BN}(C_i,C_j)
P_{\BN}(C_j,C_k)/P_{\BN}(C_j)$.  In the second case, there exists a
unique clique $C^*$ that lies on the path between any pair of these
cliques.  Again, by the separation property, $C^*$ renders these
cliques conditionally independent, so we can compute
\begin{equation}\label{eq:triples}
P_{\BN}(C_i,C_j,C_k) = \sum_{C^*} \frac{P_{\BN}(C_i,C^*) P_{\BN}(C_j,C^*)
P_{\BN}(C_k,C^*)}{P_{\BN}(C^*)^2}\,\,\text{.}
\end{equation}

Thus, we have reduced the problem to one of computing the marginals over
all pairs of cliques in a calibrated clique-tree.  We can use dynamic
programming to execute this process efficiently.  We construct a table
that contains $P_{\BN}(C_i,C_j)$ for each pair of cliques $C_i,C_j$.
We construct the table in order of length of the path from $C_i$ to $C_j$.
The base case is when $C_i$ and $C_j$ are adjacent in the tree.  In this
case, we have that
$P_{\BN}(C_i,C_j) = P_{\BN}(C_i)P_{\BN}(C_j)/P_{\BN}(C_i \cap C_j)$.
The probability expressions in the numerator are simply the clique
potentials in the calibrated tree.  The denominator can be obtained by
marginalizing either of the two cliques. In fact, this expression is
computed as a byproduct of the calibration process, so the
marginalization is not required.  For cliques $C_i$ and $C_j$ that
are not adjacent, we let $C_k$ be the node adjacent to $C_j$ on the
path from $C_i$ to $C_j$.  The clique $C_k$ is one step closer to
$C_i$, so, by construction, we have already computed $P(C_i,C_k)$.  We
can now apply the separation property again:
\begin{equation}\label{eq:pairwise}
P_{\BN}(C_i,C_j) = \sum_{C_k}
   \frac{P_{\BN}(C_i,C_k)P_{\BN}(C_k,C_j)}{P_{\BN}(C_k)}\,\,.
\end{equation}

\subsubsection{Computational Complexity}
\label{sec:maidcplex}
\begin{theorem}
The computation of $\nabla V^G(\sigma)$ can be performed in time $O(\ell^2
d^3 + u|N|d^4)$, where $\ell$ is the number of cliques in the clique tree
for $G$, $d$ is the size of the largest clique (the number of entries
in its potential), $|N|$ is the number of agents, and $u$ is the total number 
of utility nodes in the game.
\end{theorem}
\begin{proof}
The cost of calibrating the
clique tree for $\BN_\sigma$ is $O(\ell d)$.  The cost of computing
\eqref{eq:pairwise} for a single pair of cliques is $O(d^3)$, as
we must compute a factor over the variables in three cliques before
summing out.  We must perform this computation $O(\ell^2)$ times, once
for each pair of cliques, for a total cost of $O(\ell^2 d^3)$.  We now
compute marginal joint probabilities over triples of cliques $\parents_{U^i_n}$,
$\bD_n$, $\bD_{n'}$ for every utility node $U^i_n$ and every 
agent $n'$ other than $n$.  There are $u(|N|-1)$ such triples.  
Computing a factor over the variables in three cliques may first
require computing a factor over the variables in four cliques, at a
cost of $O(d^4)$.  Given this factor, computing the expected value of
the utility node takes time $O(d^3)$, which does not affect the asymptotic
running time.  The total cost for computing all the marginal joint probabilities
and expected utilities is therefore $O(u|N|d^4)$, and the total
cost for computing $\nabla V^G(\sigma)$ is $O(\ell^2 d^3 + u|N|d^4)$.   
\end{proof}

With this method, we have shown that a single iteration in the continuation
method can be accomplished in time exponential in the \emph{induced
width} of the graph --- the number of variables in the largest clique
in the clique tree.  The induced width of the optimal clique tree ---
the one with the smallest maximal clique --- is called the
\emph{treewidth} of the network.  Although finding the optimal clique
tree is, itself, an NP-hard problem, good heuristic algorithms
are known~\cite{Cowetal99}.
In games where interactions between the agents are highly structured
(the road MAID, for example), the size of the largest clique 
can be a constant even as the number of agents grows.  In
this case, the complexity of computing the Jacobian grows only
quadratically in the number of cliques, and hence also in the number
of agents.  Note that the matrix adjoint operation takes time cubic in
$m$, which is at least $|N|$,
so a single step along the path actually has cubic computational cost.

\section{Results}
We performed run-time tests of our algorithms on a wide variety of both 
graphical games and MAIDs.  Tests were performed on an Intel Xeon processor 
running at 3 GHz with 2 GB of RAM, although the memory was never taxed during 
our calculations.

\label{sec:res}

\begin{figure}
\begin{center}
\begin{tabular}{cc}
\includegraphics[viewport=37 0 542 420,width=\figwidth]{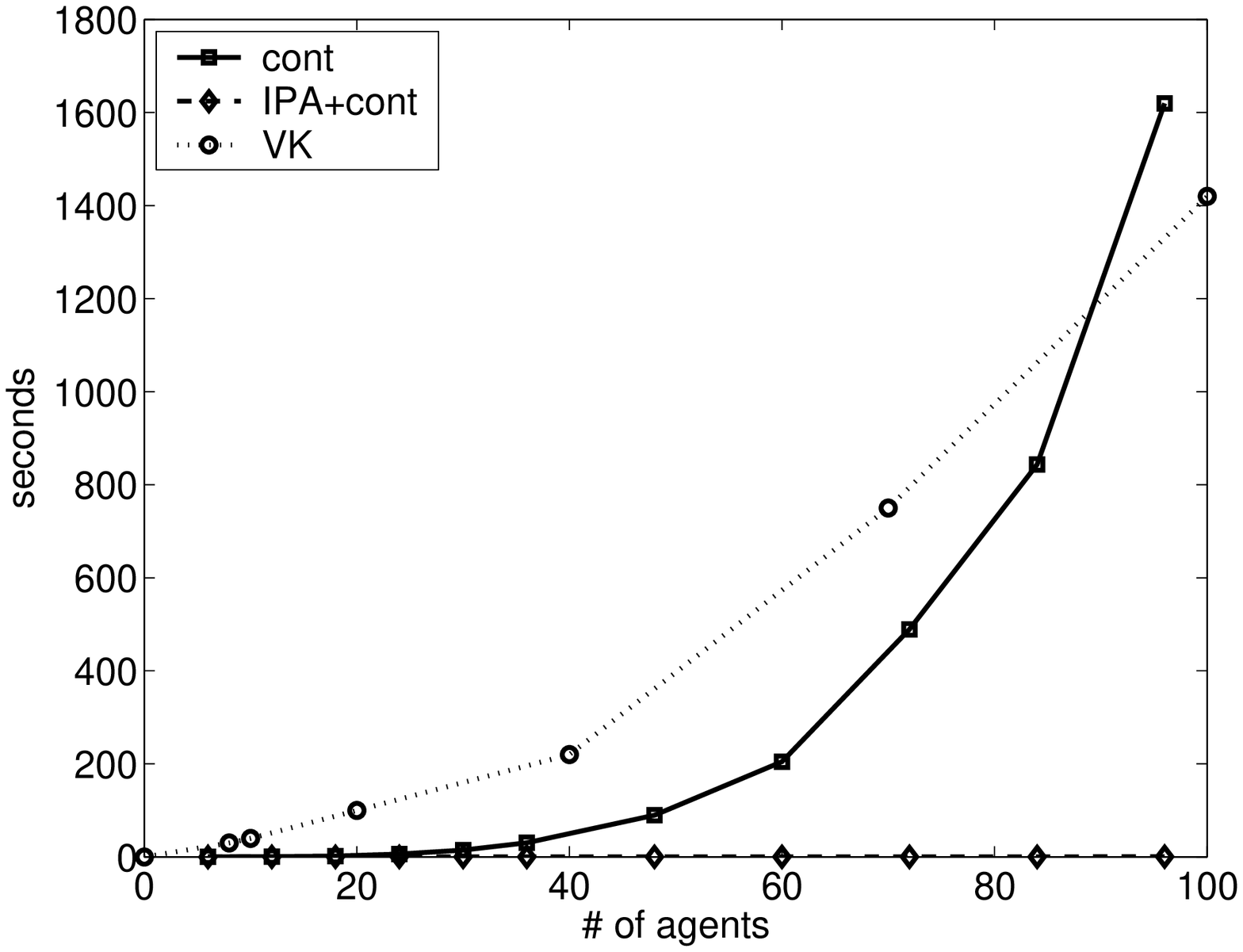}  &
\includegraphics[viewport=37 0 542 420,width=\figwidth]{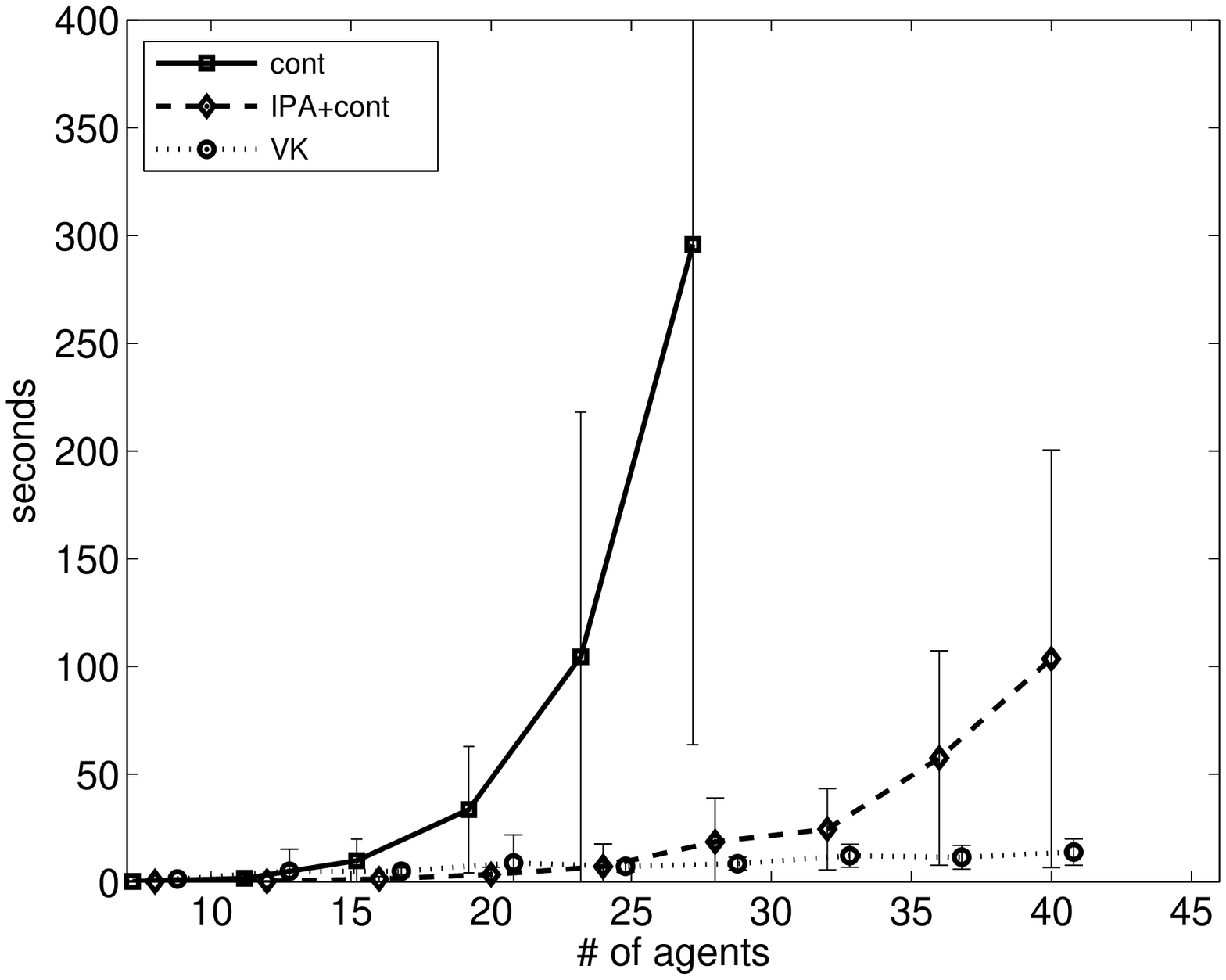}  \\
(a) & (b) \\
\includegraphics[viewport=37 0 542 440,width=\figwidth]{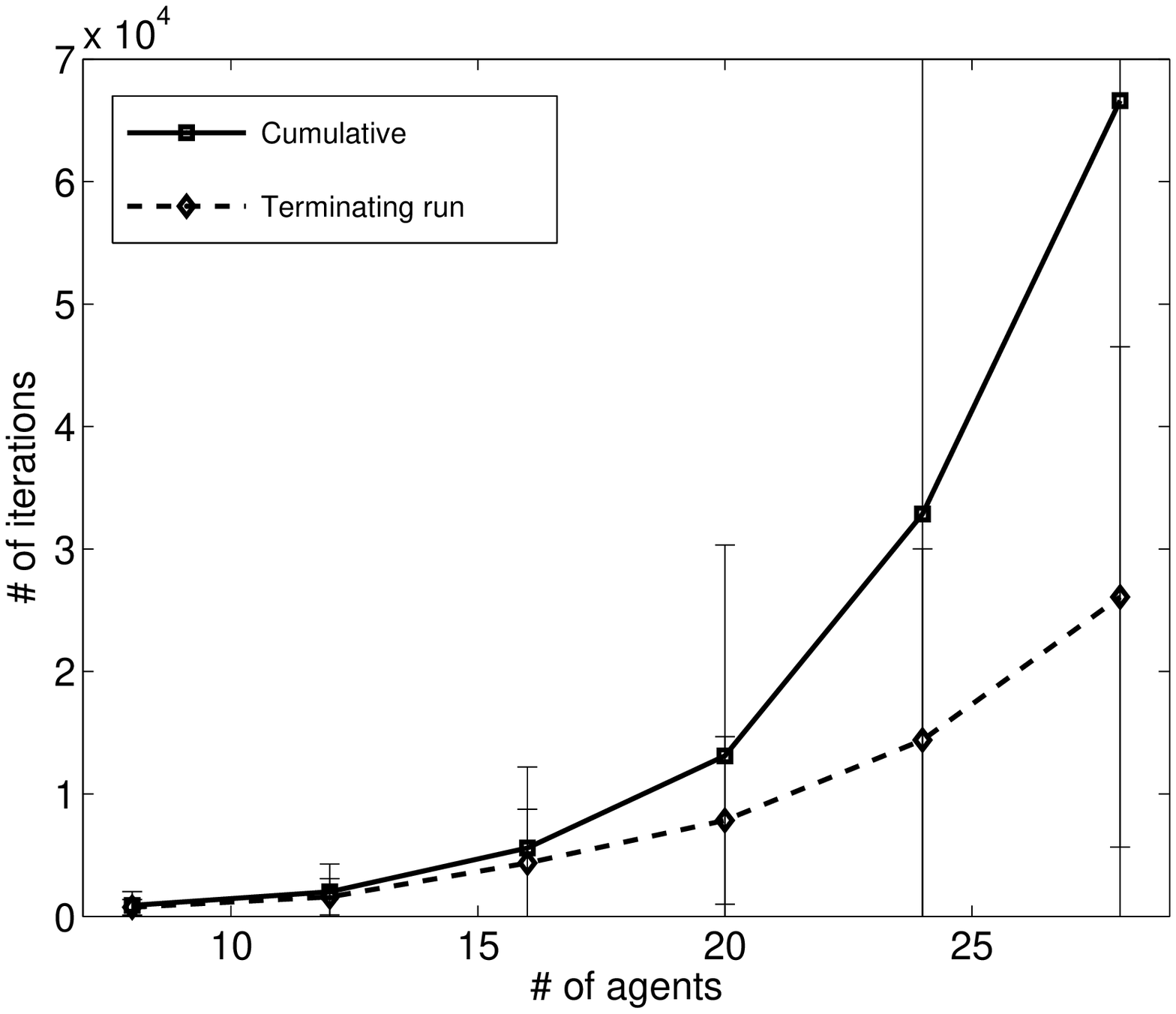} &
\includegraphics[viewport=37 0 542 440, width=\figwidth]{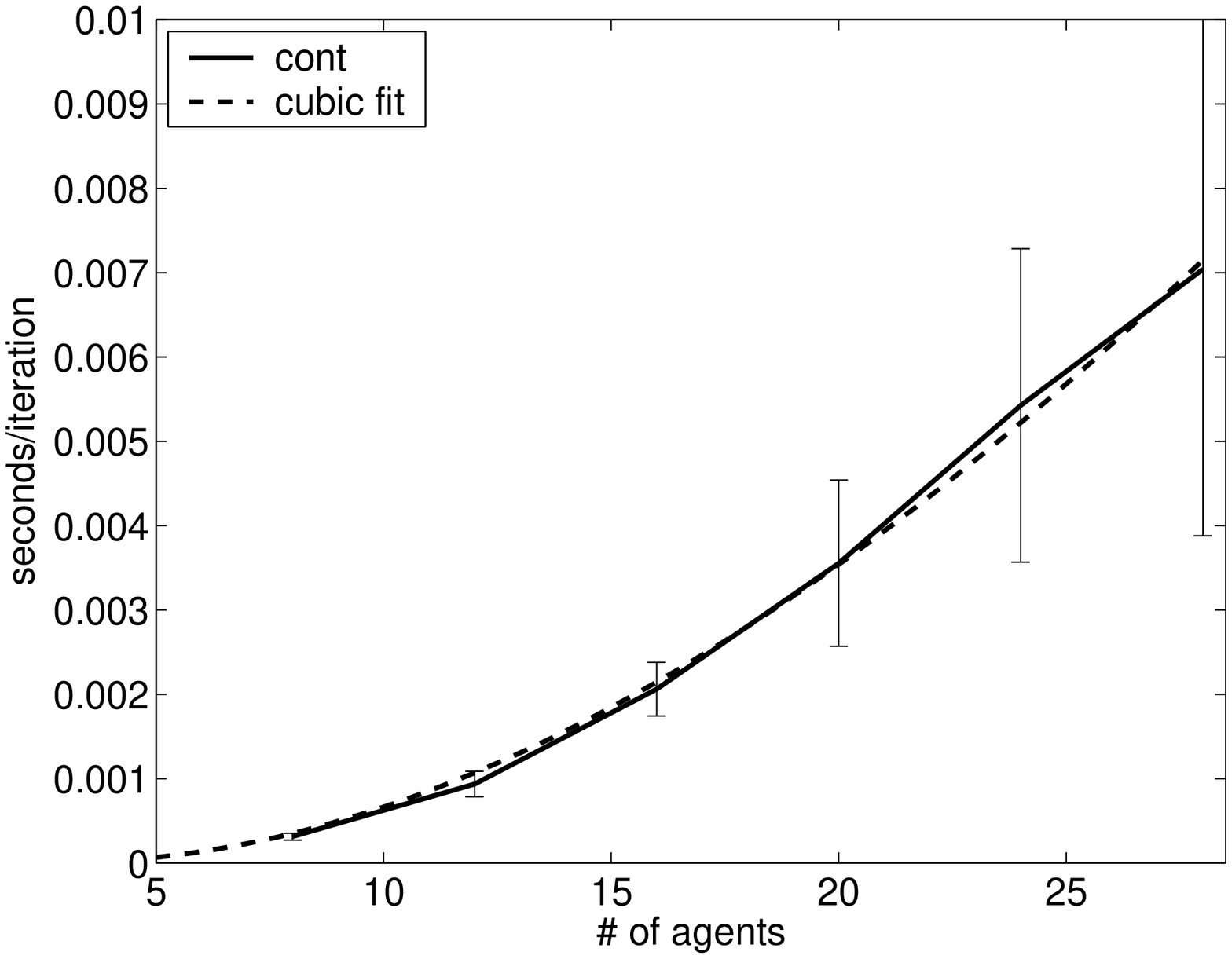} \\
(c) & (d) 
\end{tabular}
\end{center}
\caption{
Results for 2-by-$L$ road game with rock-paper-scissors payoffs:
(a) running time.
Results for road game with random payoffs:
(b) running time; (c) number of iterations of \cont;
(d) average time per iteration of \cont.
}
\label{fig:times-ggroads}
\end{figure}

\begin{figure}[h]
\begin{center}
\begin{tabular}{cc}
\includegraphics[viewport=37 0 542 420,width=\figwidth]{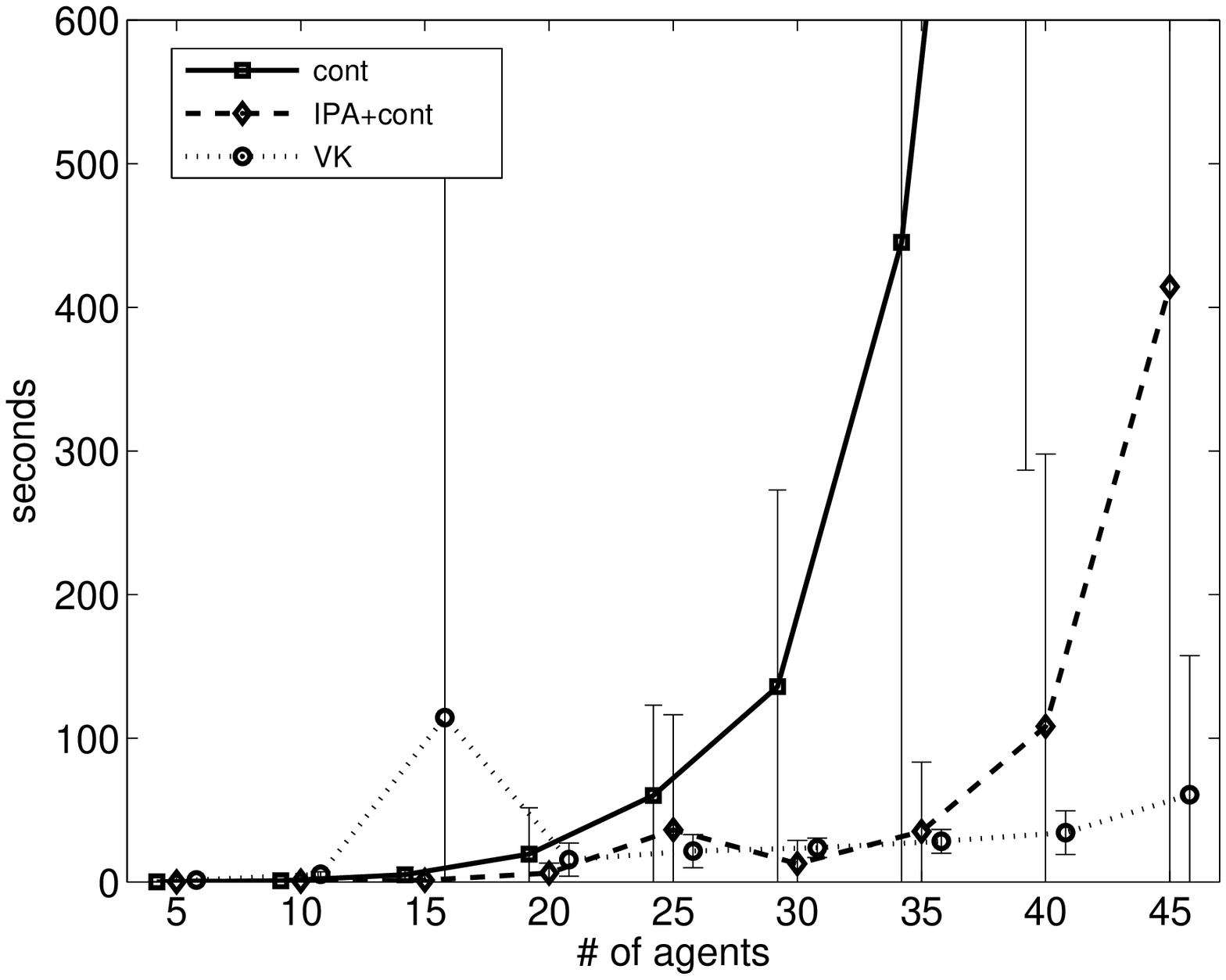}  &
\includegraphics[viewport=37 0 542 420,width=\figwidth]{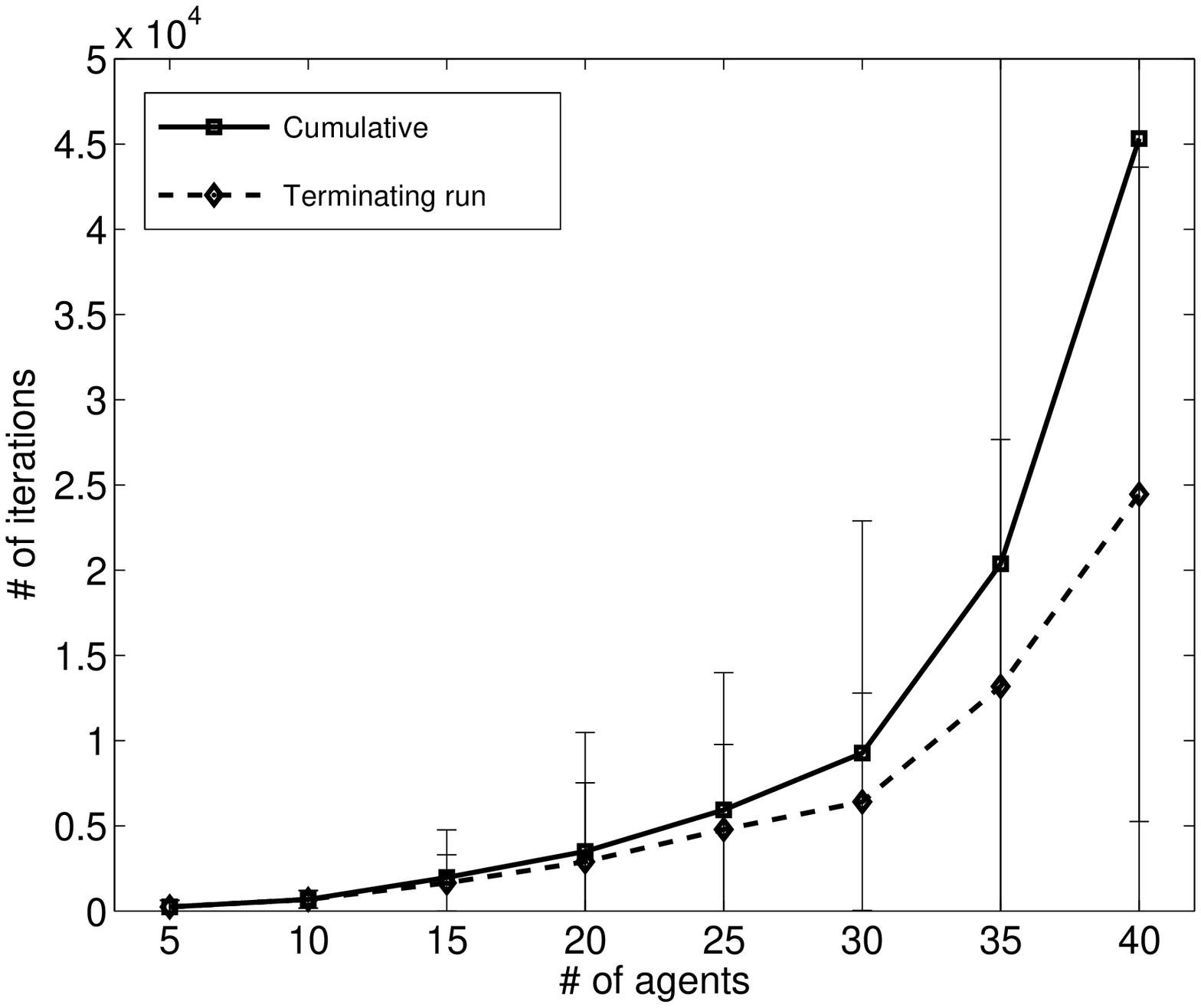} \\
(a) & (b) \\
\includegraphics[viewport=37 0 542 440,width=\figwidth]{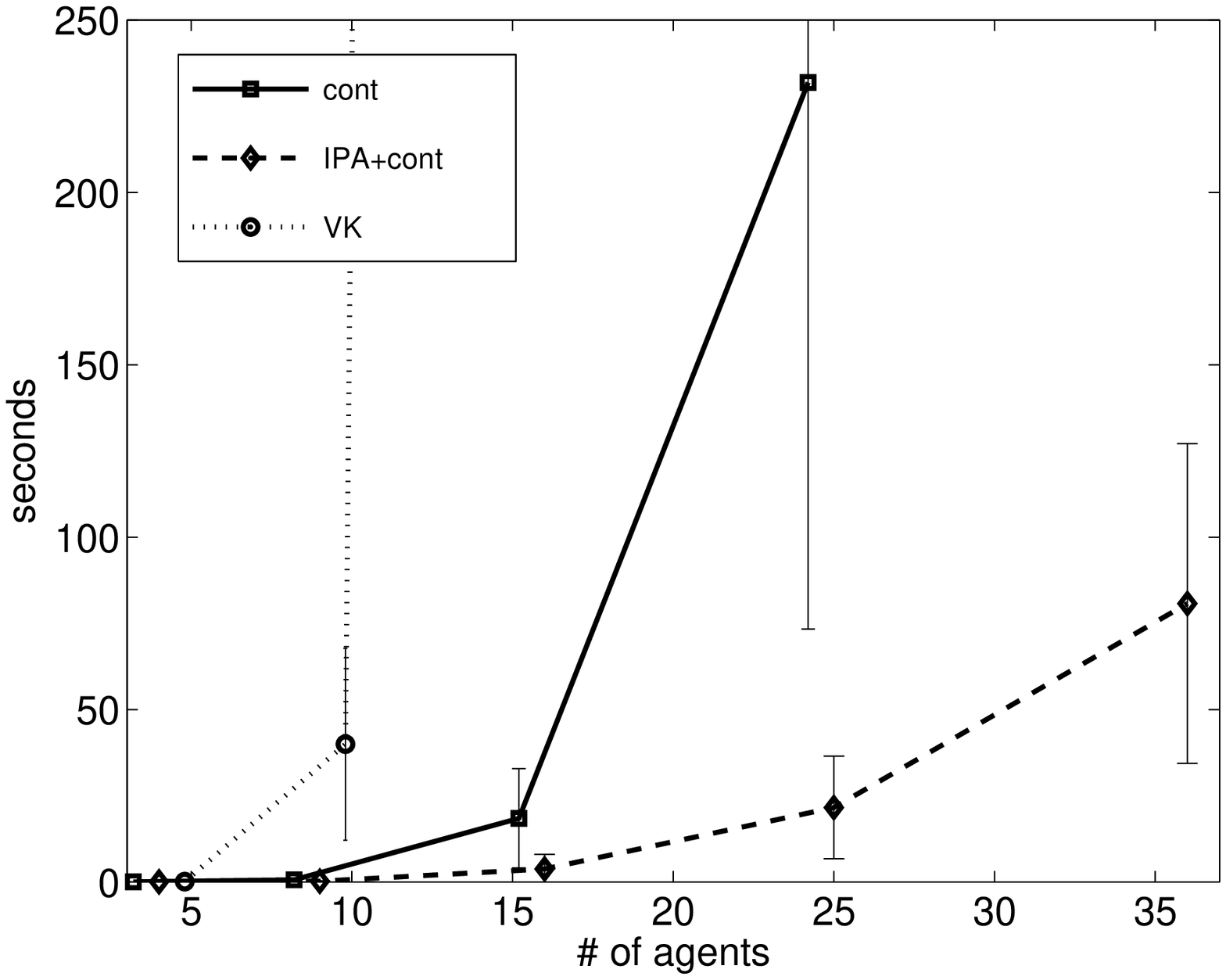} &
\includegraphics[viewport=37 0 542 440,width=\figwidth]{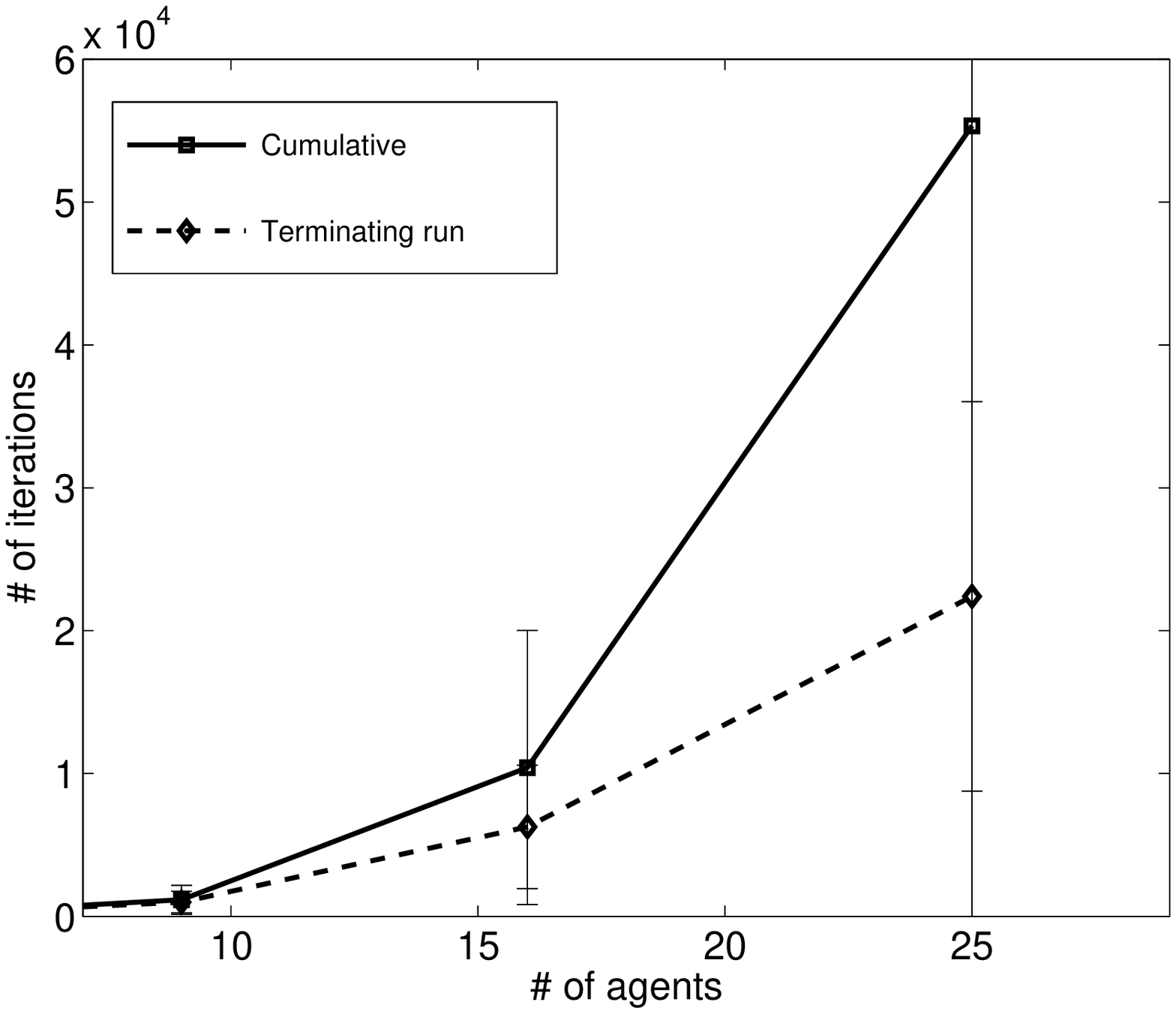} \\
(c) & (d)
\end{tabular}
\end{center}
\caption{
Results for ring game with random payoffs:
(a) running time; (b) number of iterations of \cont. 
Results for $L$-by-$L$ grid game with random payoffs:
(c) running time; (d) number of iterations of \cont.
}
\label{fig:times-ggsringgrids}
\end{figure}

\subsection{Graphical Games}

For graphical games, we compared two versions of our algorithm: \cont,
the simple continuation method, and \ipacont, the continuation method with
\ipa initialization.  We tested the hybrid equilibrium refinement
algorithm of \citeA{VicKol02} (\vk hereafter) for comparison, with
the same parameters that they used. The \vk algorithm only returns
$\epsilon$-equilibria; no exact methods exist which are comparable to
our own.

Our algorithms were run on two classes of games defined by \citeA{VicKol02} 
and two additional classes.  The road game 
of \exref{ex:roadgg}, denoting a
situation in which agents must build land plots along a road, is
played on a 2-by-$L$ grid; each agent has three actions, and its
payoffs depend only on the actions of its (grid) neighbors.  Following
\vk, we ran our algorithm on road games with additive
rock-paper-scissors payoffs: each agent's payoffs are a sum of payoffs
from independent rock-paper-scissors games with each of its neighbors.
This game is, in fact, a polymatrix game, and hence is very easy to
solve using our methods.  In order to test our algorithms on more
typical examples, we experimented with road games in which
the entries of the payoff matrix for each agent were chosen uniformly
at random from $[0,1]$.  We also experimented with a ring graph with three
actions per agent and random payoffs.  Finally, in order to test games
with increasing treewidth, we experimented with grid games with random
payoffs.  These are defined in the same manner as the road games, 
except that the game graph is an $L$-by-$L$ grid.

For each class of games, we chose a set of game sizes to run on.  For
each, we selected (randomly in cases where the payoffs were random) a
set of 20 test games to solve.  We then solved each game using \cont,
\ipacont, and \vk.  For \cont, we started with a different random
perturbation vector each time and recorded the time and number of
iterations necessary to reach the first equilibrium.  For \ipacont, we
started with a different initial strategy profile for \ipa each time and
recorded the total time for both \ipa and \cont to reach the first
equilibrium.   

All equilibria found by our algorithm had error at most $10^{-12}$, 
essentially machine precision.  The hybrid
refinement algorithm of \vk found $\epsilon$-equilibria with average error of
about $10^{-4}$ for road games with rock-paper-scissors payoffs, $0.01$
for road games and grid games with random payoffs, and $0.03$ for ring
games with random payoffs, although the equilibria had error as high as $0.05$ 
for road games and $0.1$ for ring games.  

For smaller games, the algorithms always converged to an equilibrium.
In some larger games, \cont or \ipa detected that they had entered a
cycle and terminated without finding an equilibrium.  By maintaining a
hash table of support cells they have passed through already, both \cont
and \ipa are able to detect when they have entered a support cell for the
second time.  Although this is not a sure sign that they have entered a
cycle, it is a strong indicator.  When potential cycles were detected,
the algorithms were restarted with new random initialization values.
Note that cycles in the execution of \cont can never arise if the algorithm
does not stray from the path dictated by the theory of \GW, so that random
restarts reflect a failure to  follow the path accurately.

When an equilibrium was eventually found, the cumulative time for all
the random restarts was recorded.  The error bars in the running time
graphs show the variance due to the number of random restarts required,
the choices of initialization values, and, for random games, the choice
of game.

Random restarts were required in 29\% of the games we tested.  On average,
2.2 restarts were necessary for these games.  Note that this figure is
skewed by the larger games, which occasionally required many restarts;
the largest games sometimes required 8 or 9 restarts.  In a few large
graphical games (10 random road games and 8 random ring games), \ipa
did not converge after 10 restarts; in these cases we did not record
results for \ipacont.  \cont always found an equilibrium within 10
restarts.  Our results are shown in \subfigrefs{fig:times-ggroads}{a,b,c,d}
and \subfigrefs{fig:times-ggsringgrids}{a,b,c}.

For random roads, we also plotted the number of iterations and time per
iteration for \cont in \subfigrefs{fig:times-ggroads}{c,d}.  The number
of iterations varies based both on the game and perturbation vector chosen.
However, the time per iteration is almost exactly cubic, as predicted.
We note that, when \ipa was used as a quick-start, \cont invariably
converged immediately (within a second) --- all of the time was spent in
the \ipa algorithm.

In the road games, our methods are more efficient for smaller games,
but then become more costly.  Due to the polymatrix nature of the
rock-paper-scissors road games, the \ipacont algorithm solves them
immediately with the Lemke-Howson algorithm, and is therefore significantly
less expensive than \vk.  In the random ring games, our algorithms are
more efficient than \vk for smaller games (up to 20--30 agents), with
\ipacont performing considerably better than \cont.  However, as with road
games, the running time of our algorithms grows more rapidly than that of
\vk, so that for larger games, they become impractical.  Nevertheless,
our algorithms performed well in games with up to $45$ agents and $3$
actions per agent, which were previously intractable for exact algorithms.
For the $L$-by-$L$ grid games, our algorithm performed much better than
the \vk algorithm (see \subfigrefs{fig:times-ggsringgrids}{c,d}), with
and without IPA quick-start.  This reflects the fact that the running-time complexity
of our algorithms does not depend on the treewidth of the graph.

\begin{figure}[h]
\begin{center}
\includegraphics[width=3in]{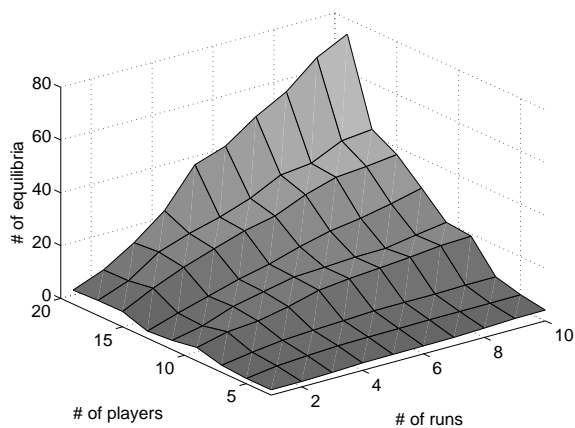}
\end{center}
\caption{The number of unique equilibria
found as a function of the size of the game and the number of runs of the
algorithm, averaged over ten random ring games.}
\label{fig:numeq-gg}
\end{figure}

We also examined the number of equilibria found by the \ipacont\
algorithm.  We ran \ipacont on the ring graphical game for differing
numbers of agents.  For each number of agents, we fixed 10 random
games, ran the algorithm 10 times on each game, and recorded the
cumulative number of unique equilibria found.  The average number 
of equilibria found over the 10 games for each number of agents is
plotted in figure~\ref{fig:numeq-gg}.  For small games (with
presumably a small number of equilibria), the number of equilibria
found quickly saturated.  For large games, there was an almost linear
increase in the number of equilibria found by each subsequent random
restart, implying that each run of the algorithm produced a new set of
solutions.

\subsection{MAIDs}
\begin{figure}
\begin{center}
\begin{tabular}{c}
\includegraphics[width=3.5in]{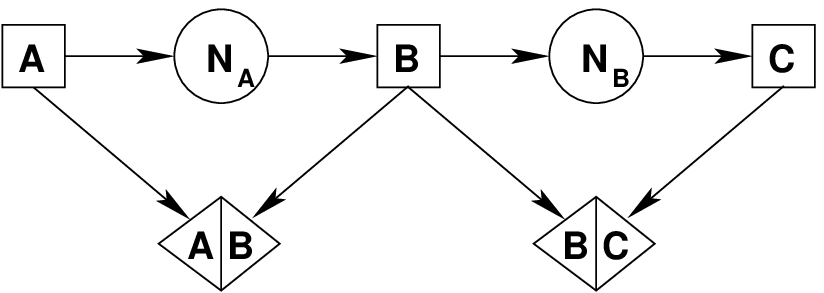} \\
(a) \\
\vspace{0.1in} \\
\includegraphics[width=3.5in]{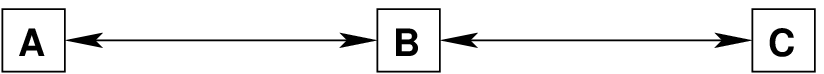} \\
(b)  
\end{tabular}
\end{center}
\caption{(a) The chain game and (b) its strategic relevance graph 
for the case of three agents (A, B, and C).}
\label{fig:chaingame}
\end{figure}

\begin{figure}[h]
\begin{center}
\begin{tabular}{cc}
\includegraphics[viewport=37 0 542 420,width=\figwidth]{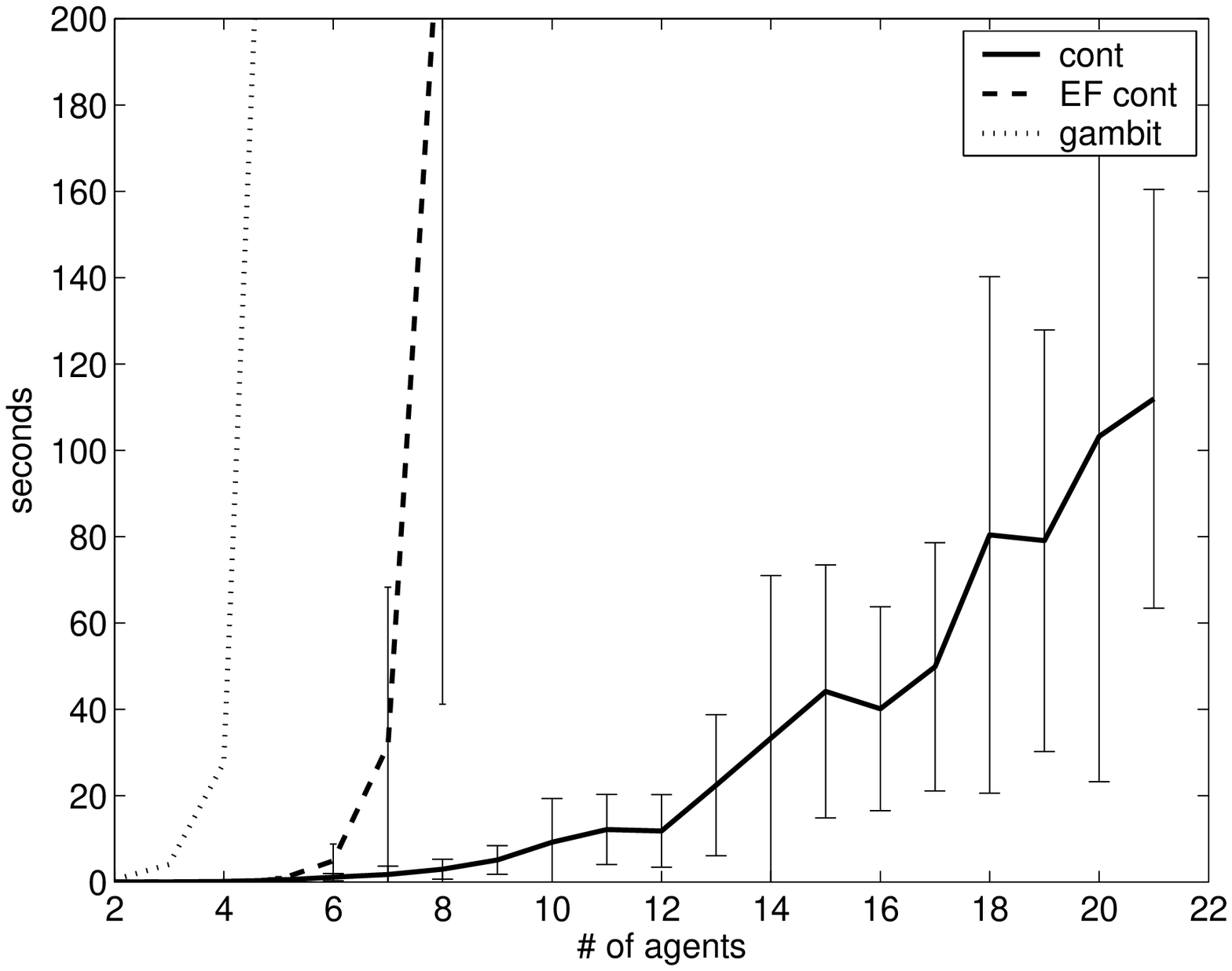} &
\includegraphics[viewport=37 0 542 420,width=\figwidth]{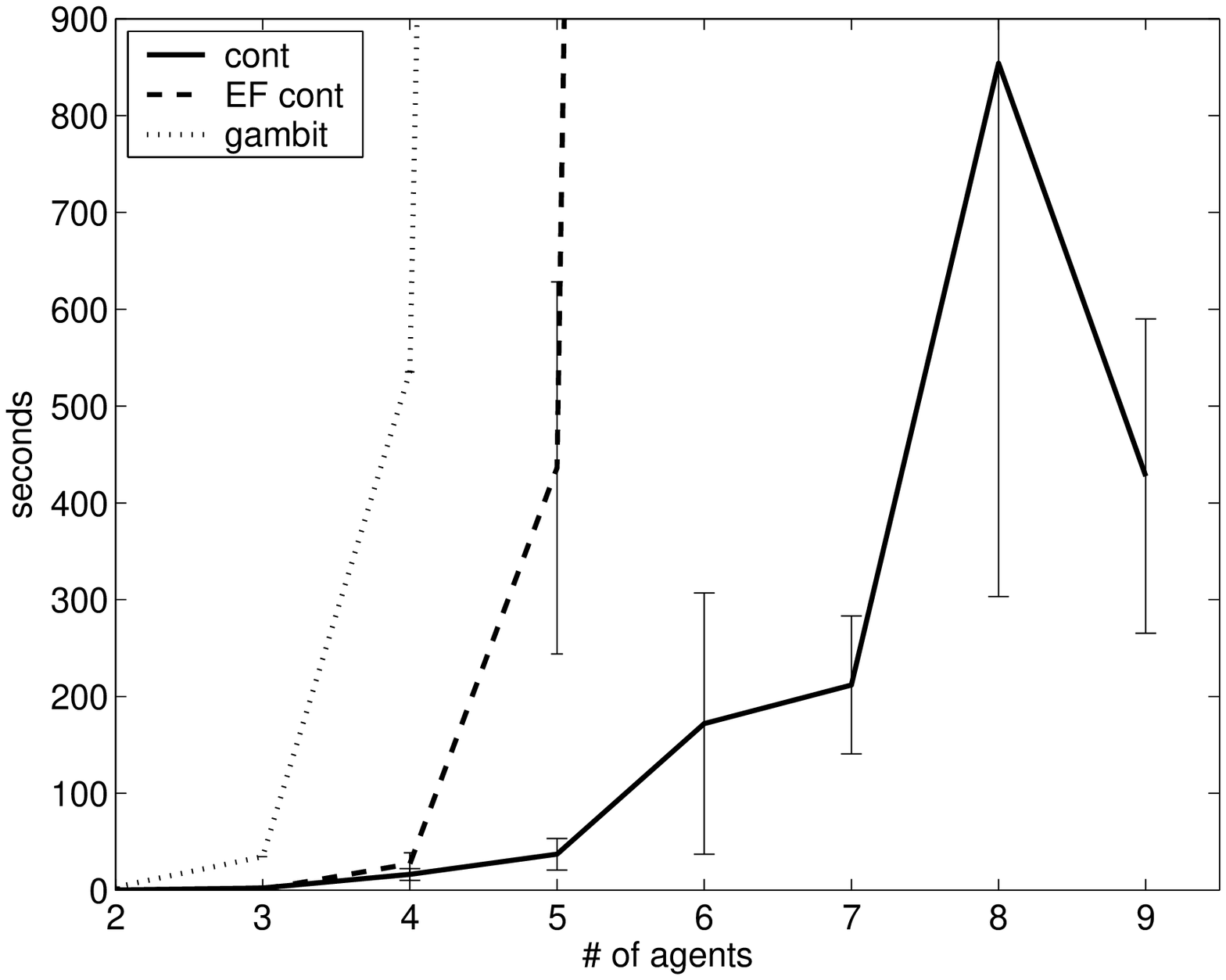} \\
(a) & (b) \\
\includegraphics[viewport=37 0 542 440,width=\figwidth]{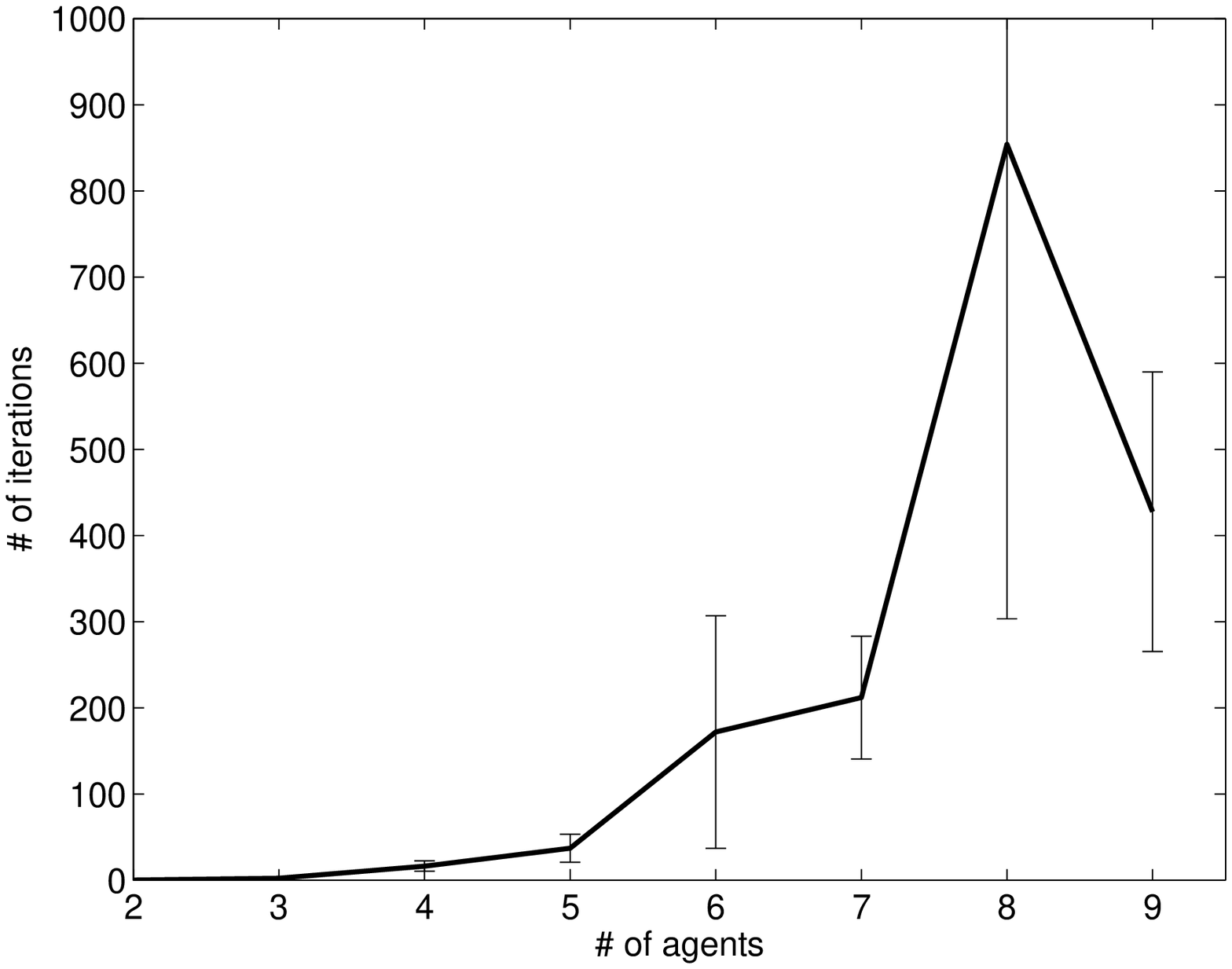} & 
\includegraphics[viewport=37 0 542 440,width=\figwidth]{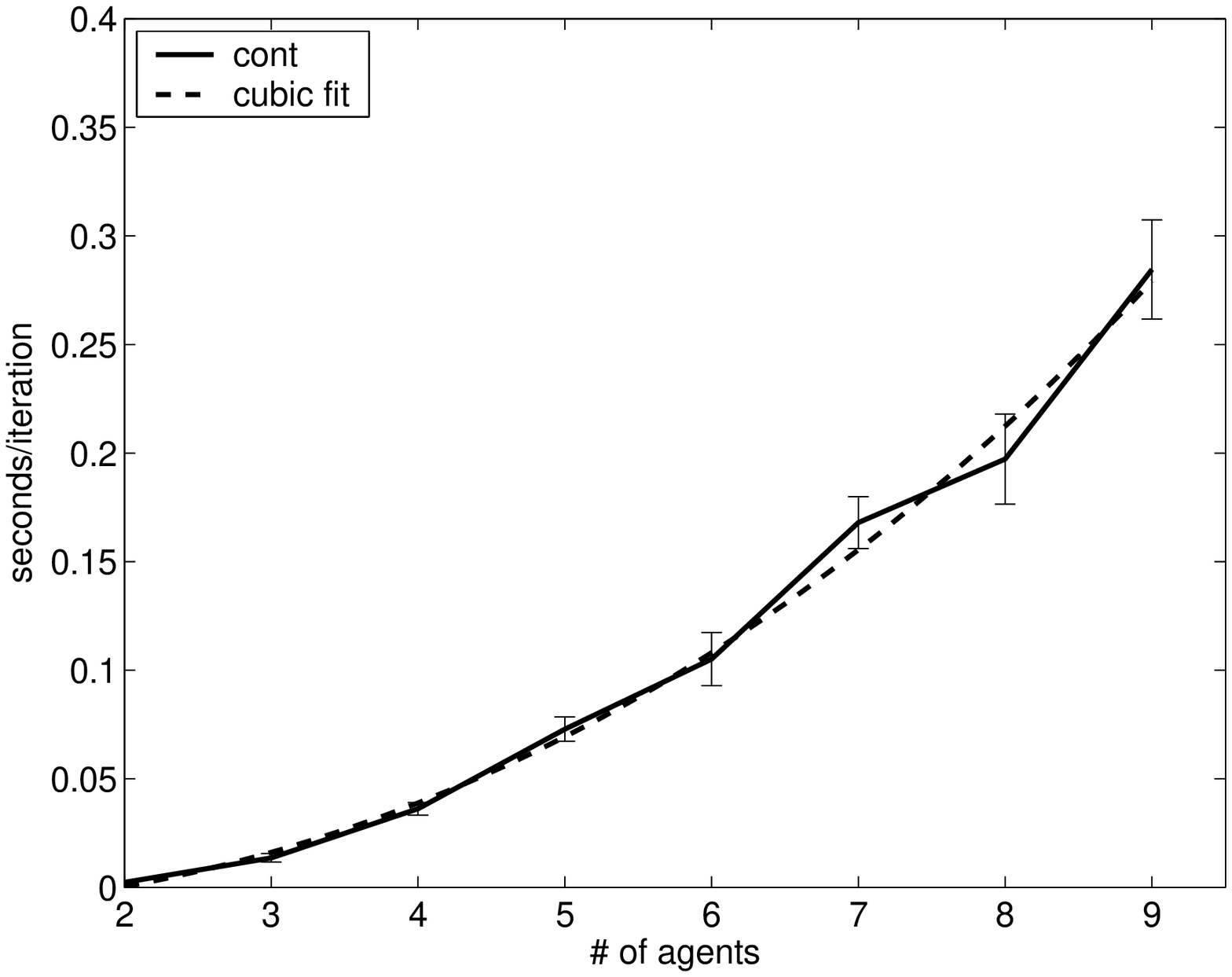} \\
(c) & (d) 
\end{tabular}
\end{center}
\caption{
Results for MAIDs:
(a) Running times for the chain MAID. 
Results for two-stage road MAID: (b) running time; (c) number of
iterations; (d) time per iteration.}
\label{fig:times-maid}
\end{figure}

The previous computational method for MAIDs~\cite{KolMil01} stopped at
strategic relevance: each SCC was converted into an equivalent
extensive-form game and solved using standard methods.  Our algorithm
takes advantage of further structure once a game has already been
decomposed according to strategic relevance.  All of our test cases
were therefore selected to have relevance graphs consisting of a single strongly
connected component.

In order to ascertain how much difference our enhancements made, we 
compared the results for our MAID algorithm, \maidcont, to those achieved by
converting the game to extensive-form and running both \efcont,
the extensive-form version of \cont as specified by \GW, and
\gambit \cite{Gambit}, a standard 
game theory software package.
The time required for conversion to extensive form is not included in
our results.

We ran our algorithms on two classes of games, with varying sizes.
The first, to which we refer as the chain game, alternates between decision and chance
nodes (see \figref{fig:chaingame}).  
Each decision node belongs to a different agent.  Each
agent has two utility nodes, each connected to its own decision node and
to a neighbor's (except for the end agents, who have one utility node
for their single neighbor).  There are three actions at each decision
node.  All probability tables and payoff matrices are chosen at
uniformly at random.  The
second class is the two-stage road building game from
\exref{ex:twostageroad}, shown in \subfigref{fig:twostageroad}{b}.
In this class, we chose payoffs carefully, by hand, to ensure non-trivial mixed strategy
equilibria.  

We ran on chain games of all sizes between 2 and 21, and road games of all sizes between
2 and 9.  For each size, we randomly selected 20 perturbation vectors and 20 games (all
20 road games were the same, since payoffs were set by hand, and all 20 chain games had payoffs
randomly assigned).  We then tested the algorithms on these games, initialized with these perturbation 
vectors, and averaged across test cases.  The timing results appear in
\subfigrefs{fig:times-maid}{a,b}.  The error bars reflect variance due to the 
choice of game (in the chain games), the choice of perturbation vector, and the number
of random restarts required.  

In some cases, as with the graphical game tests, 
\maidcont failed to find an equilibrium, terminating early because it detected that
it had entered a cycle.  In these cases, it was restarted with a new perturbation vector until
it successfully terminated.  
When an equilibrium was eventually found, the cumulative time
for all the random restarts was recorded.  Over the course of our test
runs, only two chain games required a  random restart.  Both were of
size 7.  Our algorithms failed more frequently on road games; the
spike for road games of size 8 reflects the fact that the games of
this size required, on average, 1.2 random restarts before an
equilibrium was found.  Strangely, \maidcont was much more successful
on the road game of size 9, succeeding without random restarts in all
but two cases.   

We tested \gambit and \efcont only on smaller games, because the time
and memory requirements for testing on larger ones were beyond our means.
Our results show that, while \efcont is  a faster algorithm than \gambit
for extensive-form games, it is inadequate for the larger MAIDs that we
were able to solve with \maidcont.  This is not at all surprising; a
road game of size 9 has 26 decision or chance nodes, so the equivalent
extensive-form 
game tree has $2^{26} \approx 67$ million outcome nodes.  For MAIDs
of this size, the Bayesian network inference techniques that we have used become
necessary.  

For all MAIDs, realization probabilities were constrained to be at least
$10^{-4}$ ({\it i.e.}, we found $\epsilon$-perfect equilibria with
$\epsilon=10^{-4}$).  The accuracy of these equilibria was within $10^{-12}$, 
or machine precision.

As with graphical games, we recorded the number of iterations until
convergence as well as the time per iteration for \maidcont.  The results appear
in \subfigrefs{fig:times-maid}{c,d}.  The time per iteration is fit well
by a cubic curve, in accordance with our theoretical predictions.
The variance is primarily due to the execution of the retraction operator,
whose running time depends on the number of strategies in the support.

\section{Discussion and Conclusions}
\label{sec:conc}
We have described here two adaptations of the continuation method algorithms
of \GW, for the purpose of accelerated execution on structured games.  Our results 
show that these algorithms represent significant advances
in the state of the art of equilibrium computation for both graphical games 
and MAIDs.

\subsection{Related Work on Graphical Games}
In the last few years, several papers have addressed the issue of
finding equilibria in structured games.  For graphical games, the
exact algorithms proposed so far apply only to games where the
interaction structure is an undirected tree, and where each agent has
only two possible actions.  \citeA{KeaLitSin01} provide an
exponential-time algorithm to compute all exact equilibria in such a
game, and \citeA{LitKeaSin02} provide a polynomial-time algorithm to
compute a single exact equilibrium.  
For this very limited set of
games, these algorithms may be preferable to our own, since
they come with running-time guarantees.  
However, 
it is yet to be tested whether these algorithms are, in fact, more
efficient in practice.  Moreover,
our methods are applicable to fully general games, and
our results indicate that they perform well.

More effort has been focused on the computation of
$\epsilon$-equilibria in general graphical games.  A number of 
algorithms have recently been proposed for this task.  Most of these use a
discretized space of mixed strategies: probabilities must be selected
from a grid in the simplex, which can be made arbitrarily fine.  For
computational reasons, however, this grid must typically be quite
coarse, 
as the number of grid points to consider grows exponentially with the
number of actions per agent.  Most of these methods (implicitly or explicitly) define an
equilibrium as a set of constraints over the discretized strategy
space, and then use some constraint solving method:
\citeA{KeaLitSin01} use a tree-propagation algorithm (KLS);
\citeA{VicKol02} use standard CSP variable elimination methods (VK1);
and \citeA{OrtKea03} use arc-consistency constraint propagation
followed by search (OK).  \citeA{VicKol02} also propose a gradient
ascent algorithm (VK2), and provide a hybrid
refinement method that can, with further computation, reduce the
equilibrium error.

As with the exact methods, the KLS algorithm is restricted to
tree-structured games, and comes without experimental running time
results (although it is guaranteed to run in polynomial time).
\citeA{KeaLitSin01} give a suggestion for working on a non-tree
graph by constructing the junction tree and passing messages therein.
However, the necessary computations are not clear and potentially very
expensive.

The VK1 algorithm is applicable to graphical games of arbitrary
topology, with any number of actions per agent.  It takes time
exponential in the treewidth of the graph.  If the treewidth is
constant, then it scales linearly with the number of agents; however,
our results show that it very quickly becomes infeasible if the
treewidth expands (as in the grid game).

Both of these methods come with complexity guarantees, which depend on
the treewidth of the graph.  The others (OK and VK2, as well as our
algorithm) are insensitive to treewidth --- a single iteration takes
time polynomial in the size of the game representation (and hence
exponential only in the maximum degree of the graph).  However, they
all require an unknown number of iterations to converge.
Corollary~\ref{cor:gridnp} shows that, in general, computation of
equilibria with discretized strategies in games with fixed degree is
hard.  Thus, the lack of complexity guarantees for these methods is
not surprising. 

Nonetheless, experimental results for OK seem promising --- they
indicate that, on average, relatively few iterations are required for
convergence.  Results indicate that OK is capable of solving grid
games of at least 100 agents (although in these cases $\epsilon$ was
as large as $0.2$, not much better than in a random fully mixed
strategy profile).  However, no running time results are provided.

VK2 also exhibits strong experimental results.  \citeA{VicKol02}
have successfully found $\epsilon$-equilibria in games of up to 400
agents, with errors of up to 2\% of the maximal payoff.  

The main drawback to these algorithms is that they only compute $\epsilon$-equilibria.
An $\epsilon$-equilibrium may be sufficient
for certain applications: if the utility functions
are themselves approximate, an agent certainly might be satisfied with
an $\epsilon$-best response; and if we make the assumption
that it is slightly costly for agents to change their minds, each
agent might need an incentive greater than $\epsilon$ to deviate.
However, $\epsilon$-equilibria do bring their own set of problems.  
The primary one is that there is no guarantee of an exact equilibrium 
in the neighborhood of an $\epsilon$-equilibrium.  This can make it
very difficult to find $\epsilon$-equilibria with small values of $\epsilon$;
attempts to refine a given $\epsilon$-equilibrium may fail.  
The lack of a nearby Nash equilibrium also implies a certain instability.  If some agent is unsatisfied with the $\epsilon$-equilibrium, play may
deviate quite far from it.
Finally, $\epsilon$-equilibria are more numerous than Nash equilibria
(uncountably so, in general).
This exacerbates the difficulty an agent faces in choosing
which equilibrium to play.

The algorithms for computing $\epsilon$-equilibria are frequently faster than our own, especially when the approximations are crude or the games have more than 50 or so agents.  However, the exact equilibria found by our algorithms are more satisfying solutions, and our results show that the performance of our algorithm is
comparable to that of approximate methods in most cases.  Surprisingly,
for many games, running time results show that ours is the fastest
available, particularly in the case of games with large treewidth, such as
the grid game in our test cases.  
Furthermore, since we can use any approximate equilibrium
as a starting point for our algorithm, advances in approximate methods
complement our own method.  The hybrid algorithm
of~\citeA{VicKol02} turns out to be unsuited to this purpose,
as it tends not to remove any pure strategies from the support, but
it is interesting to see whether other methods 
(including those listed above) 
might be more effective.
It remains to be seen how small $\epsilon$ must be for our methods to
reliably refine an approximate equilibrium.

\subsection{Related Work on MAIDs}
\citeA{KolMil01} (KM) define a notion of dependence between agents'
decisions (s-relevance), and provide an algorithm that can decompose
and solve MAIDs based on this fairly coarse independence structure.
Our algorithm is able to exploit finer-grained structure, resolving an
open problem left by KM.  
In general, our method will not automatically exploit the same
structure obtained by decomposing the game into its relevance
components, and so our methods are best regarded as a complement to
those of KM; after decomposition according to s-relevance, our
algorithm can be applied to find equilibria efficiently in the
decomposed problems.  Running time results indicate that our methods
are significantly faster than previous standard algorithms for
extensive-form games.  This is unsurprising, since the game
representation of our test cases is 
exponentially larger in the number of players when converted to
extensive-form.

\citeA{Vic02} proposes an approximate hill-climbing algorithm for
MAIDs that takes advantage of the same sort of fine-grained structure
that we do: Bayesian network inference is employed to calculate
expected utility as one component of the score function for a single
iteration.  A constraint-satisfaction approach is also proposed.
However, these proposals were never implemented, so it is hard to
determine what quality equilibria they would find or how quickly they
would find them.  

\citeA{LaM00} proposes a continuation method for finding one or
all equilibria in a G net, a representation that is very similar to
MAIDs.  This proposal only exploits a very limited set of structural
properties (a strict subset of those exploited by KM).  
This proposal was also never implemented, and several
issues regarding non-converging paths seem unresolved.

Our algorithm is therefore the first to be able to exploit the
finer-grained structure of a MAID.  Moreover, our algorithm, applied in
conjunction with the decomposition method of KM, is able to take 
advantage of the full known independence structure in a MAID.
A potential drawback is the requirement that strategies be
$\epsilon$-perturbed.  However, decreasing $\epsilon$ incurs no
additional computational cost, although there are limits imposed by
machine precision.   
Perfect equilibria --- a highly desirable refinement
of Nash equilibria, defined to be the limit of a sequence of 
$\epsilon$-perturbed equilibria as $\epsilon$ goes to zero --- can therefore be 
computed effectively by our algorithm
with little or no additional computational cost.  
In this sense, our use of perturbed
strategies is advantageous.  We have not implemented a local search
algorithm to find an exact perfect equilibrium in the neighborhood of
a found $\epsilon$-perturbed equilibrium, although it should be
straightforward to do so.

\subsection{Conclusion and Further Work}
We have presented two related algorithms for computing exact equilibria in
structured games.  Our algorithms are based on the methods of \GW, but
perform the key computational steps in their methods 
much more efficiently by exploiting game structure.  Our
approach yields the first exact algorithm to take advantage of
structure in general graphical games and the first algorithm to take full advantage of
the independence structure of a MAID.  These algorithms are capable of
computing exact equilibria in games with large numbers of agents, which were
previously intractable for exact methods.  

Our algorithms come without theoretical running time bounds, but we have
noticed certain interesting trends.  In both the graphical game and the
MAID version of our algorithm, each iteration executes in time polynomial
in the number of agents, so we have examined the number of iterations
required for convergence.  
Our adaptive step size technique decreases the number of random
restarts required to find an equilibrium, but increases
the number of iterations required to cross a support cell in larger
games.  When adaptive step size is disabled, we 
have noticed that the number of iterations required, averaged across games 
with random payoffs, seems to grow approximately linearly.  
Intuitively, it makes sense that the number of iterations should
be at least linear: starting from a pure strategy profile, a linear number
of actions (in the number of agents) must enter the support in order for
us to reach a general strategy profile.  Each support boundary requires at
least one iteration of our algorithm.  It is somewhat surprising, however,
that the number of iterations required does not grow more quickly.  It is
an interesting open problem to analyze the number of iterations required
for convergence.

In very large games, the tendency of our algorithm to cycle
increases.  This phenomenon can be attributed, partially, to the cumulative
effect of ``wobbling'': after a great number of wobbles, it is
possible that the path has been altered sufficiently that it does not
pass through an equilibrium.  We have noticed that some games seem
intrinsically harder than others, requiring many random restarts
before convergence.  For very large games, the overall
running time of our algorithm is therefore quite unpredictable.  

Our algorithms might be improved in a number of ways.  Most importantly,
the continuation method would profit greatly from more sophisticated path-following methods;
in a number of cases, \cont or \maidcont failed to find an equilibrium because
it strayed too far from the path.  Better path-following techniques might
greatly increase the reliability of our algorithms, particularly if they
obviated the need for ``wobbles,'' which negate \GW's theoretical guarantee
of the convergence of the continuation method.  

There are also a number of theoretical questions about the algorithms of \GW
that remain unresolved.  Nothing is known about the worst-case or average-case
running time of \ipa, and no theoretical bounds exist on the number of iterations
required by \cont.  It is interesting to speculate on how the choice of 
perturbation ray might affect the execution of the algorithm.  
Can the algorithm be directed toward particular
equilibria of interest either by a careful selection of the perturbation ray
or by some change in the continuation method?  Is there a way of selecting
perturbation rays such that all equilibria will be found?  Is there a way
of selecting the perturbation ray so as to speed up the execution time?

Several improvements might be made to \maidcont.  We have not adapted \ipa
for use in MAIDs, but it should be possible to do so, making use of the
generalized Lemke algorithm of \citeA{KolMeg96} to solve intermediate 
linearized MAIDs.  
The computation of $\nabla V^G$ might also be accelerated using a variant
of the all-pairs clique tree algorithm that only computes the potentials
for pairs of \emph{sepsets} --- sets of variables shared by adjacent cliques ---
rather than pairs of cliques.

Our work suggests several interesting avenues for further research.
In fact, after the initial publication of these results \cite{BluSheKol03}, at least one further application
of our techniques has already been developed:  
\citeA{BhaLey04} have shown that an adaptation of \cont can be used
to efficiently solve a new class of structured games called action-graph games \cite<a generalization of local effect games as presented in>{LeyTen03}.  We believe
that these games, and other structured representations, show great promise as enablers
of new applications for game theory.  They have several advantages over their 
unstructured counterparts: they are well-suited to games with
a large number of agents, they are determined by fewer parameters, making it 
feasible for human researchers to fully
specify them in a meaningful way, and their built-in structure makes them a more 
intuitive medium in which to frame structured, real-world scenarios.  However,
to avoid the computational intractability of the general problem,
each new class of structured games requires a new algorithm for equilibrium
computation.  We hypothesize that \cont and \ipa are an excellent starting point
for addressing this need.  

\smallskip
\noindent{\bf Acknowledgments.} 
This work was supported by ONR MURI 
Grant N00014-00-1-0637, and by Air Force contract F30602-00-2-0598 under
DARPA's TASK program. 
Special thanks to Robert Wilson, for kindly taking 
the time to guide us through the details of his work with Srihari Govindan, 
and to David Vickrey, for aiding us in testing our algorithms alongside his.  
We also thank the anonymous referees for their helpful comments.
\appendix
\section{Table of Notation}
\label{sec:tablenot}
\begin{center}
\begin{tabular}{|c|p{5in}|}
\hline
\multicolumn{2}{|l|}{\it Notation for all games} \\
\hline
$N$ & set of agents \\
\hline
$\sigma_n$ & strategy for agent $n$ \\
\hline
$\Sigma_n$ & strategy space for agent $n$ \\
\hline
$\sigma$ & strategy profile \\
\hline
$\Sigma$ & space of strategy profiles \\
\hline
$\sigma_{-n}$ & strategy profile $\sigma$ restricted to agents other than $n$ \\
\hline
$\Sigma_{-n}$ & space of strategy profiles for all agents other than $n$ \\
\hline
$(\sigma_n,\sigma_{-n})$ & strategy profile in which agent $n$ plays strategy $\sigma_n$ and all other agents act according to $\sigma_{-n}$ \\
\hline
$G_n(\sigma)$ & expected payoff to agent $n$ under strategy profile $\sigma$ \\
\hline
$V^G(\sigma)$ & vector deviation function \\
\hline
$R$ & retraction operator mapping points to closest valid strategy profile \\
\hline
$F$ & continuation method objective function \\
\hline
$\lambda$ & scale factor for perturbation in continuation method \\
\hline
$\bw$ & free variable in continuation method \\
\hline
\multicolumn{2}{|l|}{\it Notation for normal-form games} \\
\hline
$a_n$ & action for agent $n$ \\
\hline
$A_n$ & set of available actions for agent $n$ \\
\hline
$\ba$ & action profile \\
\hline
$A$ & set of action profiles\\
\hline
$\ba_{-n}$ & action profile $\ba$ restricted to agents other than $n$ \\
\hline
$A_{-n}$ & space of action profiles for agents other than $n$ \\
\hline
\multicolumn{2}{|l|}{\it Notation for extensive-form games} \\
\hline
$z$ & leaf node in game tree (outcome) \\
\hline
$Z$ & set of outcomes \\
\hline
$i$ & information set \\
\hline
$I_n$ & set of information sets for agent $n$ \\
\hline
$A(i)$ & set of actions available at information set $i$ \\
\hline
$H_n(y)$ & sequence (history) for agent $n$ determined by node $y$ \\
\hline
$Z_h$ & set of outcomes consistent with sequence (history) $h$ \\
\hline
$b(a|i)$ & probability under behavior profile $b$ that agent $n$ will choose action $a$ at $i$ \\
\hline
$\sigma_n(z)$ & realization probability of outcome $z$ for agent $n$ \\
\hline
\multicolumn{2}{|l|}{\it Notation for graphical games} \\
\hline
$\family_n$ & set of agent $n$ and agent $n$'s parents \\
\hline
$\Sigma_{-n}^f$ & strategy profiles of agents in $\family_n$ other than $n$ \\
\hline
$A_{-n}^f$ & space of action profiles of agents in $\family_n$ other than $n$ \\
\hline
\multicolumn{2}{|l|}{\it Notation for MAIDs} \\
\hline
$D_n^i$ & decision node with index $i$ belonging to agent $n$ \\
\hline
$U_n^i$ & utility node with index $i$ belonging to agent $n$ \\
\hline
$\parents_X$ & parents of node $X$ \\
\hline
$\dom(S)$ & joint domain of variables in set $S$ \\
\hline
\end{tabular}
\end{center}
\section{Proof of \thmref{thmnphard}}
\label{nphardproof}
\begin{proof}
The proof is by reduction from 3SAT.  For a given 3SAT instance, we
construct a graphical game whose equilibria encode satisfying
assignments to all the variables.

Let $\clauses=\{c_1,c_2,\ldots,c_m\}$ be the clauses of the 3SAT
instance in question, and let $\vars = \left\{v_1, \lnot v_1, v_2, \lnot v_2, \ldots, v_n,
\lnot v_n\right\}$ be the set of literals.  If a
variable appears in only one clause, it can immediately be assigned so
as to satisfy that clause; therefore, we assume that variables
appear in at least two clauses.

We now construct the (undirected) graphical game.  For each clause,
$c_i$, we create an agent $C_i$ connected to $C_{i-1}$ and $C_{i+1}$
(except $C_1$ and $C_m$, which only have one clause neighbor).  We also create agents $V_i^\ell$ for
each literal $\ell$ in $c_i$ (there are at most 3).  If, for example,
$c_i$ is the clause $(\lnot v_1 \lor v_2)$, it has agents
$V_i^{\lnot v_1}$ and $V_i^{v_2}$.  We connect each of these to $C_i$.
For every variable $v$, we group all agents $V_i^{v}$ and $V_j^{\lnot
v}$ and connect them in a line, the same way we connected clauses to
each other.  The order is unimportant. 

\begin{figure}
\begin{center}
\includegraphics[width=3in]{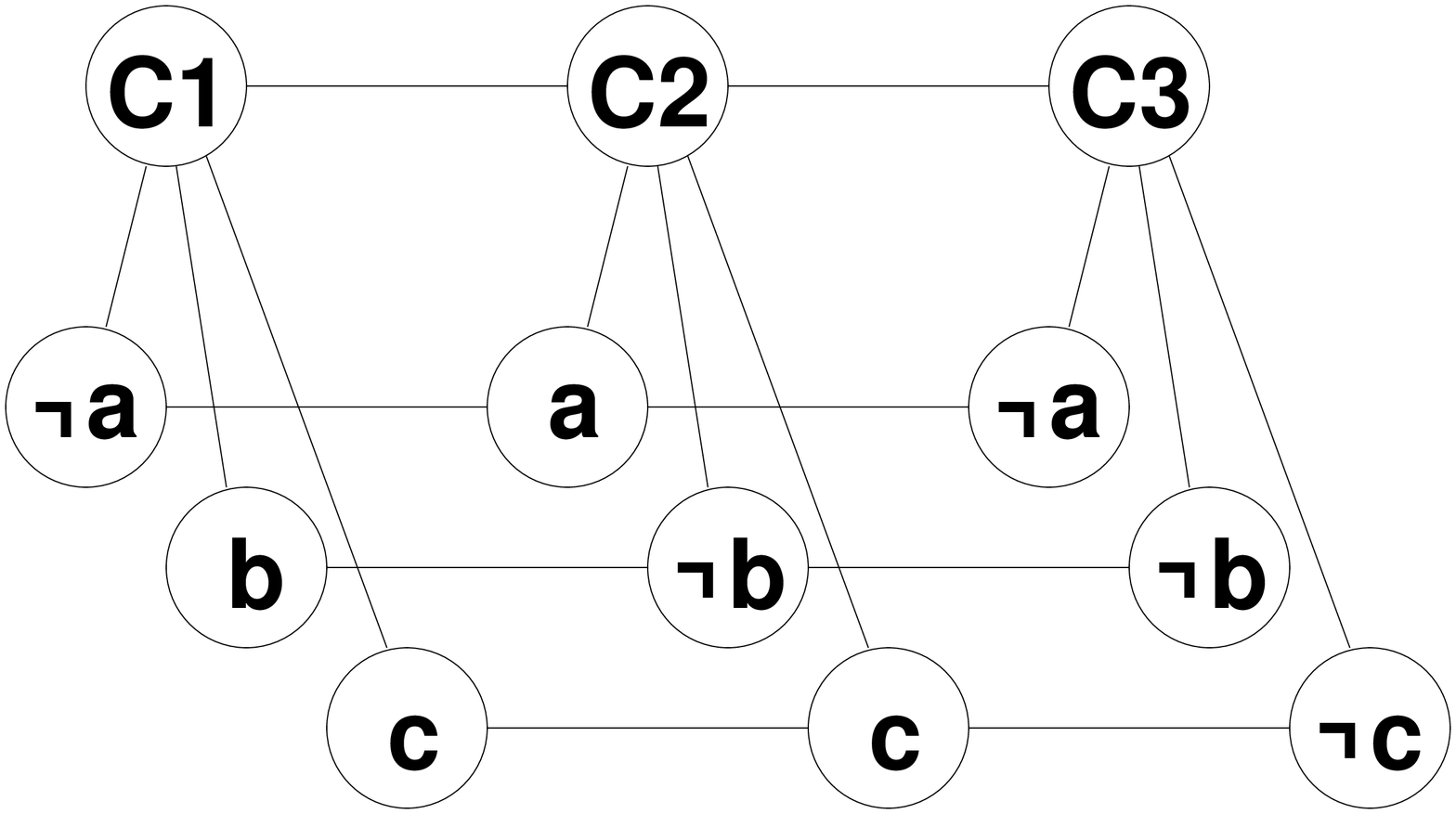}
\end{center}
\caption{ 
Reduction of the 3SAT instance $(\lnot a \lor
b \lor c) \land (a \lor \lnot b \lor c) \land (\lnot a \lor \lnot b
\lor \lnot c)$ to a graphical game. }
\label{fig:3SAT}
\end{figure}

Clause agents now have at most $5$ neighbors (two clauses on either
side of them and three literals) and literal agents have at most $3$
neighbors (two literals on either side of them and one clause).  This
completely specifies the game topology.  As an example,
\figref{fig:3SAT} shows the graphical game corresponding to the 3SAT
problem $(\lnot a \lor b \lor c) \land (a \lor \lnot b \lor c) \land
(\lnot a \lor \lnot b \lor \lnot c)$.

Now we define the actions and payoff structure.  Each agent can be interpreted as a Boolean variable, and has two
actions,~\true~and~\false, which correspond to the Boolean values true and
false.  Intuitively, if a clause $C_i$ plays~\true, it is satisfied.
If an agent $V_i^v$ plays~\true, where $v$ is a non-negated variable, 
then $v$ is assigned to be true.  If $V_j^{\lnot v}$ plays~\true, then 
$v$ is assigned to be false.  

The payoff matrix for a clause agent $C_i$ is designed to ensure that
if one clause is unsatisfied, the entire 3SAT instance is marked
as unsatisfied.  It can best be expressed in
pseudo-code, as follows:

\begin{algorithmic}
\IF{any of $C_i$'s clause neighbors play \false}
\STATE payoff is $ 
\left\{ \begin{array}{l}
1\text{ \quad for playing \false} \\
0\text{ \quad for playing \true}
\end{array} \right.$
\ELSIF{at least one of $C_i$'s literals plays \true~($C_i$ is satisfied)}
\STATE payoff is $
\left\{ \begin{array}{l}
2\text{ \quad for playing \false} \\
2\text{ \quad for playing \true}
\end{array} \right.$
\ELSE
\STATE ($C_i$ is unsatisfied)
\STATE payoff is $ 
\left\{ \begin{array}{l}
1\text{ \quad for playing \false} \\
0\text{ \quad for playing \true}
\end{array} \right.$
\ENDIF
\end{algorithmic}

The payoff matrix for a literal agent $V_i^\ell$ is designed to 
encourage agreement with the other literals along the line for the variable $v(\ell)$ 
associated with $\ell$.  It can be described in pseudo-code as follows:

\begin{algorithmic}
\IF{the parent clause $C_i$ plays \false}
\STATE payoff is $ 
\left\{ \begin{array}{l}
1\text{ \quad for playing consistently with a false assignment to $v(\ell)$} \\
0\text{ \quad for playing the opposite}
\end{array} \right.$
\ELSIF{$V_i^\ell$'s literal neighbors all play consistently with a single assignment to $v(\ell)$}
\STATE payoff is $
\left\{ \begin{array}{l}
2\text{ \quad for playing consistently with neighbors} \\
0\text{ \quad for playing the opposite}
\end{array} \right.$
\ELSE
\STATE payoff is $ 
\left\{ \begin{array}{l}
2\text{ \quad for playing consistently with a false assignment to $v(\ell)$} \\
0\text{ \quad for playing the opposite}
\end{array} \right.$
\ENDIF
\end{algorithmic}

If the formula does have a satisfying assignment, then there is a pure
equilibrium in which each literal is consistent with the assignment
and all clauses play~\true; in fact, all agents receive higher
payoffs in this case than in any other equilibrium, so that
satisfying assignments correspond to equilibria with maximum social
welfare.  

If the parent clauses all play~\false, then clearly at equilibrium all
non-negated literals must play~\false~and all negated literals must
play~\true.  This is the trivial equilibrium.  It remains to
show that the trivial equilibrium is the only equilibrium for
unsatisfiable formulas, \textit{i.e.} that any non-trivial equilibrium
can be used to construct a
satisfying assignment.  We first prove two simple claims.

\begin{claim}
\label{clm:clause-same}
In any Nash equilibrium, either all clauses play \true with probability one
or all clauses play \false with probability one.
\end{claim}
\begin{proof}
In no case is it advantageous for a clause to
choose~\true~over~\false, and if a neighbor clause takes the
action~\false, it is in fact disadvantageous to do so.  Thus, if any
clause has a non-zero probability of playing~\false~at an equilibrium,
its neighbors, and consequently all other clauses, must play~\false~with 
probability one.  Therefore,
 the only possible equilibria have 
all clauses playing~\false~or all clauses playing~\true.  
\end{proof}

It follows immediately from this claim that every non-trivial equilibrium
has all clauses playing $\true$ with probability one.  

\begin{claim}
\label{clm:var-same}
In any non-trivial Nash equilibrium, in a line of literals for the same variable $v$,
all those literals that play pure strategies must choose them consistently 
with a single assignment to $v$. 
\end{claim}
\begin{proof}
Since the equilibrium is non-trivial, all clauses play $\true$. 
Suppose that one of the literals, $V^\ell$, employs the pure strategy
corresponding to a 
false assignment to $v$.  It suffices to show that in fact all literals
in the line must have pure strategies corresponding to a false assignment to $v$.  
Consider a neighbor $V^{\ell'}$ of $V^\ell$.  Either $V^{\ell'}$'s neighbors (one of which is $V^\ell$)
 both play consistently with a false assignment to $v$, 
 in which case $V^\ell$ must also play consistently
with a false assignment to $v$, or its neighbors play inconsistently, in which case
the $\bf{else}$ clause of $V^\ell$'s payoff matrix applies and $V^{\ell'}$ must, again, 
play consistently with a false assignment to $v$.  We may proceed
all the way through the line in this manner.  All literals in the line must 
therefore have pure strategies consistent
with a false assignment to $v$, so there can be no contradicting literals.
\end{proof}

Suppose we have a non-trivial equilibrium.  Then by \clmref{clm:clause-same},
all clauses must play \true with probability $1$.  If all of the literals
have pure strategies, it is clear that the equilibrium corresponds to
a satisfying assignment: the literals must all be consistent with an
assignment by \clmref{clm:var-same}, and the clauses
must all be satisfied.  Some subtleties arise when we consider mixed
strategy equilibria.

Note first that in each clause, the payoff for choosing~\true~is the
same as for choosing~\false~in the case of a satisfying assignment to
its literals, and is less in the case of an unsatisfying assignment.  Therefore, if
there is any unsatisfying assignment with non-zero probability, the
clause must play~\false.

Consider a single clause $C_i$, assumed to be choosing~\true~at
equilibrium.  The mixed strategies of $C_i$'s literals induce a
distribution over their joint actions.  Because $C_i$ plays~\true,
each joint action with non-zero probability must satisfy $\bigvee_\ell
V_i^\ell$.  If a literal $V_i^\ell$ has a mixed strategy, consider what will
happen if we change its strategy to either one of the possible pure
strategies (\true~or~\false).  Some of the joint actions with non-zero
probability will be removed, but the ones that remain will be a subset
of the originals, so will still satisfy $\bigvee_\ell V_i^\ell$.
Essentially, the value of $\ell$ does not affect the satisfiability of
$C_i$, so it can be assigned arbitrarily.

Thus, if each literal in a line for a certain variable has a mixed
strategy, we can assign the variable to be either true or false (and
give each literal in the line the corresponding pure strategy) without
making any of the clauses connected to these literals unsatisfied.  In
fact, we can do this if all literals in a line that have pure
strategies are consistent with each other: if there are indeed
literals with pure strategies, we assign the variable according to
them.  And by \clmref{clm:var-same}, this will always be the case. 
\end{proof}

We observe briefly that this constructed graphical game has only a
finite number of equilibria, even if peculiarities in the 3SAT instance
give rise to equilibria with mixed strategies.  If all clauses 
play~\false, then there
is only one equilibrium.  If all clauses play~\true, then we can
remove them from the graph and trim the payoff matrices of the
literals accordingly.  Each line of literals is in this case a generic
graphical game, with a finite set of equilibria.  The equilibria of the original
game must be a subset of the direct product of these finite sets.  

\bibliography{bblum06a}
\bibliographystyle{theapa}

\end{document}